\documentclass{article}

\usepackage{arxiv}

\usepackage[utf8]{inputenc} 
\usepackage[T1]{fontenc}    
\usepackage{hyperref}       
\usepackage{url}            
\usepackage{booktabs}       
\usepackage{amsfonts}       
\usepackage{nicefrac}       
\usepackage{microtype}      
\usepackage{lipsum}		

\DeclareMathOperator*{\argmin}{arg\,min}

\title{Data-Driven Predictive Modeling of Neuronal Dynamics using Long Short-Term Memory}

\author{
  Benjamin ~Plaster \\
  Department of Chemical and Materials Engineering\\
  University of Idaho\\
  Moscow, ID 83844 \\
  \texttt{plas2554@vandals.uidaho.edu} \\
   \And
 Gautam.~Kumar \\
  Department of Chemical and Materials Engineering\\
  University of Idaho\\
  Moscow, ID 83844 \\
  \texttt{gkumar@uidaho.edu} \\
}

\begin{document}
\maketitle

\begin{abstract}
Modeling brain dynamics to better understand and control complex behaviors underlying various cognitive brain functions are of interests to engineers, mathematicians, and physicists from the last several decades. With a motivation of developing computationally efficient models of brain dynamics to use in designing control-theoretic neurostimulation strategies, we have developed a novel data-driven approach in a long short-term memory (LSTM) neural network architecture to predict the temporal dynamics of complex systems over an extended long time-horizon in future. In contrast to recent LSTM-based dynamical modeling approaches that make use of multi-layer perceptrons or linear combination layers as output layers, our architecture uses a single fully connected output layer and reversed-order sequence-to-sequence mapping to improve short time-horizon prediction accuracy and to make multi-timestep predictions of dynamical behaviors. We demonstrate the efficacy of our approach in reconstructing the regular spiking to bursting dynamics exhibited by an experimentally-validated 9-dimensional Hodgkin-Huxley model of hippocampal CA1 pyramidal neurons. Through simulations, we show that our LSTM neural network can predict the multi-time scale temporal dynamics underlying various spiking patterns with reasonable accuracy. Moreover, our results show that the predictions improve with increasing predictive time-horizon in the multi-timestep deep LSTM neural network.
\end{abstract}


\section{Introduction}
\noindent Our brain generates highly complex nonlinear responses at multiple temporal scales, ranging from few milliseconds to several days, in response to external stimulus \cite{salmelin1994dynamics,fox2005human,kiebel2008hierarchy}. One of the long-time interests in computational neuroscience is to understand the dynamics underlying various cognitive and non-cognitive brain functions by developing computationally efficient modeling and analysis approaches. In the last four decades or so, several advancements have been made in the direction of dynamical modeling and analysis of brain dynamics \cite{gerstner2014neuronal,siettos2016multiscale,breakspear2017dynamic}. In the context of modeling the dynamics of single neurons, several modeling approaches, ranging from detailed mechanism-based biophysiological modeling to simplified phenomenological/probabilistic modeling, have been developed to understand the diverse firing patterns (e.g., simple spiking to bursting) observed in electrophysiological experiments \cite{herz2006modeling,gerstner2009good}. These models provide a detailed understanding of various ionic mechanisms that contribute to generate specific spiking patterns as well as allow to perform large-scale simulations to understand the dynamics underlying cognitive behaviors. However, most of these models are computationally expensive from the perspective of developing novel real-time neurostimulation strategies for controlling neuronal dynamics at single neurons and network levels. In this paper, we investigate purely data-driven long short-term memory (LSTM) based recurrent neural network (RNN) architectures in multi-timestep predictions of single neuron’s dynamics for the use in developing novel neurostimulation strategies in an optimal control framework. 

Availability of an abundant amount of data and advances in machine learning have recently revolutionized the field of predictive data-driven dynamical modeling of complex systems using neural networks (NNs) and deep learning approaches.  Various nonlinear system identification approaches have been developed to map static input-output relations using multi-layer perceptrons (MLPs) \cite{chen1992neural}, \cite{purwar2007nonlinear}, \cite{kuschewski1993application}, \cite{pan2018long} and their variations \cite{gupta1997modeling}, \cite{patra1999identification}. Reinforcement learning has recently been explored in robotics dynamical modeling in \cite{nagabandi2018neural}. NN architectures that make use of vanilla recurrent neural network (RNNs) elements have also been explored for nonlinear system identification and modeling in \cite{jaeger2004harnessing}, \cite{bailer1998recurrent}, \cite{lenz2015deepmpc}.  However, network architectures that make use of vanilla recurrent layers often suffer from the exploding or vanishing gradient problem when used to model dynamics over long time series horizons \cite{pascanu2013difficulty}.  In \cite{lenz2015deepmpc}, a highly specialized multi-phase training algorithm was used to ensure that the network did not suffer from this problem.  LSTM based approaches to modeling dynamical systems \cite{mohajerin2019multistep}, \cite{lin2018deep}, \cite{gonzalez2018non} mitigate the vanishing gradient problem but suffer from poor early trajectory predictive performance when using long predictive horizons \cite{wang2017new}, \cite{mohajerin2019multistep}.  Additionally, LSTMs have been used to model high-dimensional chaotic systems \cite{vlachas2018data}, but these studies have been limited to single step prediction applications.

\begin{figure}[htp]
	\centering
	\includegraphics[width=\textwidth]{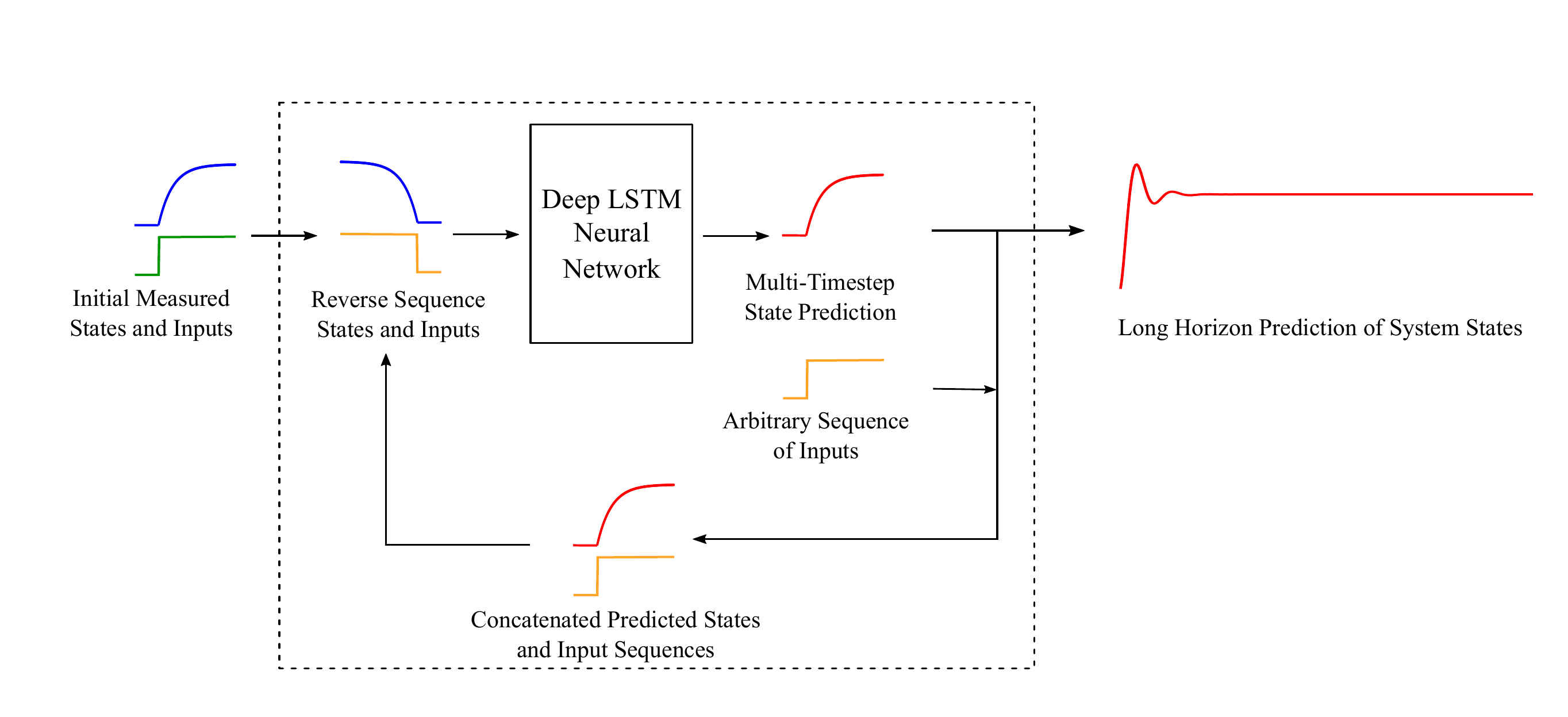}
	\caption{A schematic illustrating the overall data-driven approach developed in this paper for multi-timestep predictions of high-dimensional dynamical systems' behavior over a long time-horizon. An initial sequence of states and inputs are fed to the ``Stacked LSTM Network'' in a reverse-order for multi-timestep prediction of the system's states (``Reverse-order sequence-to-sequence mapping''). The predicted output from each stacked LSTM network is concatenated with the next sequence of inputs and fed into the next stacked LSTM network in a reverse-order to increase the predictive horizon. This process is iterated an arbitrary number of times, creating long dynamical predictions.}
	\label{fig:general_approach}
\end{figure}

In this paper, we have developed a novel deep LSTM neural network architecture, which can make multi-timestep predictions in large-scale dynamical systems. Figure \ref{fig:general_approach} shows our overall approach.  

\noindent In contrast to existing approaches in modeling dynamical systems using neural networks, our architecture uses (1) stacked LSTM layers in conjunction with a single densely connected layer to capture temporal dynamic features as well as input/output features, (2) sequence-to-sequence mapping, which enables multi-timestep predictions, and (3) reverse ordered input and measured state trajectories to the network, resulting in highly accurate early predictions and improved performance over long horizons. We show the efficacy of our developed approach in making stable multi-timestep predictions of various firing patterns exhibited by hippocampal CA1 pyramidal neurons, obtained from simulating an experimentally validated highly nonlinear 9-dimensional Hodgkin-Huxley model of CA1 pyramidal cell dynamics, over long time-horizons.  

The remaining paper is organized as follows. In Section \ref{Method}, we describe our developed deep LSTM neural network architecture and methodological approach to data-driven multi-timestep predictions of dynamical systems. We show the efficacy of our approach in making stable multi-timestep predictions over long time-horizons of neuronal dynamics in Section \ref{results} which is followed by a thorough discussion on the limitations of our approach in Section \ref{discussion}.

\section{Neural Network Architecture, Algorithm and Approach}\label{Method}

\noindent In Section \ref{LSTM_NN}, we describe our developed deep LSTM neural network architecture which combines stacked LSTMs with a fully-connected dense output layer. We describe the sequence-to-sequence mapping with reversed order input sequences used in this paper in Section \ref{SS_NN}. In Section \ref{syndata}, we provide the details on the synthetic data used to train our networks. Finally, in Section \ref{training_NN}, we provide the details on the approach used to train the developed neural network architecture.

\subsection{Deep LSTM Neural Network Architecture}\label{LSTM_NN}

\begin{figure}[htp]
	\centering
	\includegraphics[scale=0.4]{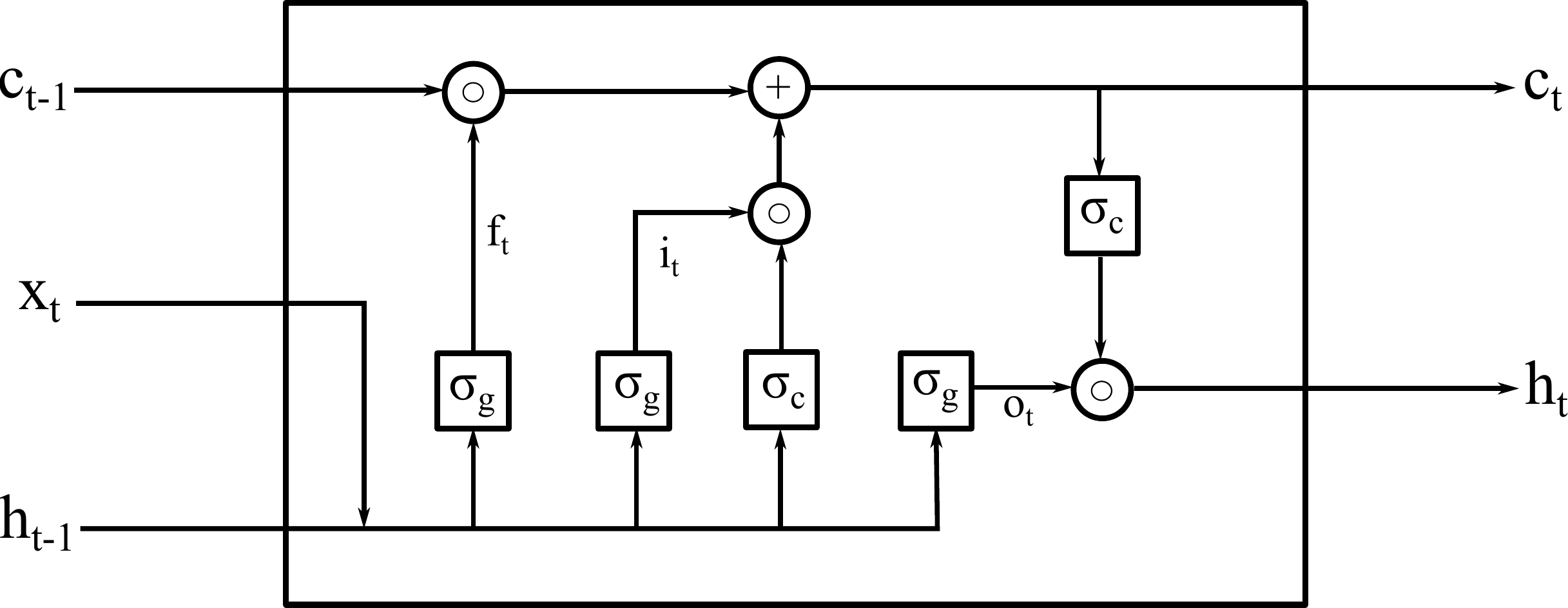}
	\caption{A schematic illustrating the internal gating operation in a single LSTM cell. The "+" represents an additive operation and the "$\circ$" represents a multiplicative operation. $\sigma_g$ is the sigmoidal activation function and $\sigma_c$ is the hyperbolic tangent activation function.}
	\label{fig:LSTM_cell}
\end{figure}

\noindent Long short-term memory (LSTM) neural networks \cite{hochreiter1997long} are a particular type of recurrent neural networks (RNNs) which mitigate the vanishing or exploding gradient problem during the network training while capturing both the long-term and the short-term temporal features in sequential time-series data processing \cite{pascanu2013difficulty}.  Specifically, LSTM uses multiple gating variables that control the flow of information of a hidden cell state and assign temporal importance to the dynamical features that are present in the time series data flowing through the cell state. Figure \ref{fig:LSTM_cell} shows a schematic illustrating the internal gating operation in a single LSTM cell.

\noindent A forward pass of information through a single LSTM cell is described by the following cell and gating state equations (reference):

\begin{subequations}
	\begin{equation}\label{gating1}
	c_t = f_t \circ c_{t-1} + i_t \circ \sigma_c(W_c x_t + U_c h_{t-1} + b_c),
	\end{equation}
	\begin{equation}\label{gating2}
	h_t = o_t \circ \sigma_c(c_t).
	\end{equation}
	\begin{equation}\label{gating3}
	f_t = \sigma_g(W_f x_t + U_f h_{t-1} + b_f), 
	\end{equation}
	\begin{equation}\label{gating4}
	i_t = \sigma_g(W_i x_t + U_i h_{t-1} + b_i),
	\end{equation}
	\begin{equation}\label{gating5}
	o_t = \sigma_g(W_o x_t + U_o h_{t-1} + b_o),
	\end{equation}		
\end{subequations}

\noindent In equations ~\eqref{gating1}-~\eqref{gating5}, $c_t \in {\rm I\!R}^h$ and  $h_t \in {\rm I\!R}^h$  represent the cell state vector and the hidden state vector, respectively, at time $t$. $f_t \in {\rm I\!R}^h$,  $i_t \in {\rm I\!R}^h$, and $o_t \in {\rm I\!R}^h$ are the ``forget gate’’, ``input gate’’, and ``output gate’’ activation vector, respectively, at time $t$. $x_t \in {\rm I\!R}^d$ is the input vector to the LSTM unit at time $t$, and $h_{t-1} \in {\rm I\!R}^h$ is the previous time step hidden state vector passed back into the LSTM unit at time $t$. The matrices $W_f$, $W_i$, and $W_o$ represent the input weights for the ``forget gate’’, ``input gate’’, and ``output gate’’, respectively. The matrices $U_f$, $U_i$ and $U_o$ represent the weights of the recurrent connections for the ``forget gate’’, ``input gate’’, and ``output gate’’, respectively. The vectors $b_f$ $b_i$, and $b_o$ represent the ``forget gate’’, ``input gate’’, and ``output gate’’ biases, respectively. $\circ$ represents the element-wise multiplication. The function $\sigma_g$ represents the sigmoidal activation function, and $\sigma_c$ is the hyperbolic tangent activation functions.

In this paper, we use stacked LSTM network integrated with a fully connected feedforward output layer to make multi-timestep state predictions. The use of a single feedforward dense output layer allows the network to effectively learn the static input-output features, while the stacked LSTM network captures the temporal dynamical features. To appropriately select the optimum dimensionality of the hidden states in a single hidden layer, we systematically varied the number of hidden states in a sequence of \{$n$, $n^2$, $2n^2$, $4n^2$, $\cdots$\}, where $n$ is the dimension of the system's state and evaluated the training performance for each case.  We found that for our application ($n = 9$), a hidden state dimensionality of $4n^2=324$ was optimal in learning dynamical behaviors while avoiding overfitting. To select the number of hidden layers, we systematically increased the number of hidden layers of identical hidden state dimensionality (i.e., 324 states) and compared the network performance during the training. We found that increasing the number of hidden layers beyond 3 layers did not improve the network performance on the training and validation dataset. Thus, we fixed the number of hidden layers to 3 in our study. Throughout this paper, we utilized stateless LSTMs which reset the internal cell and hidden states to zero after processing and performing gradient descent for a given minibatch. We initialized the network weights using the Xavier method \cite{glorot2010understanding}. Specifically, the initial weights were drawn from a uniform distribution using
\begin{equation}
W_{ij} \sim \mathcal{U}\bigg(-\frac{6}{\sqrt{n_j + n_{j+1}}},\frac{6}{\sqrt{n_j + n_{j+1}}}\bigg),
\end{equation}
where $n_j$ is the dimensionality of the input units in the weight tensor, and $n_{j+1}$ is the dimensionality of the output units in the weight tensor.

\begin{figure}[htp]
	\centering
	\includegraphics[width=\textwidth]{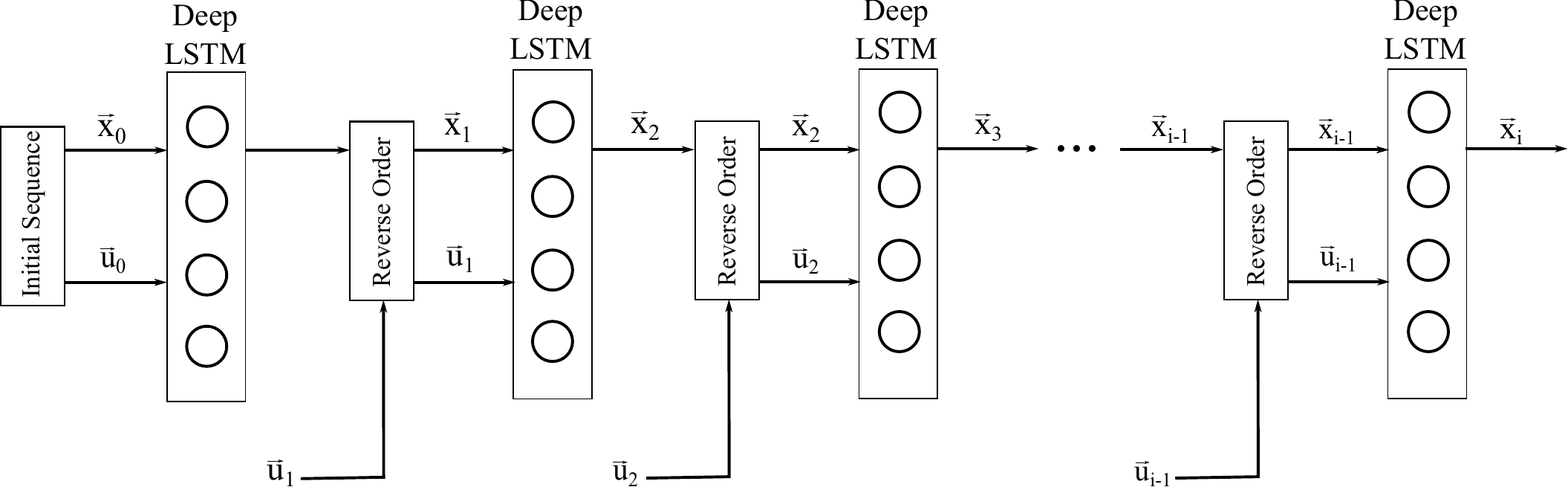}
	\caption{Iterative prediction of the system's outputs over a long time-horizon.  Each "Deep LSTM" receives the predicted sequence of outputs from the previous "Deep LSTM" and an equivalent length of new system's inputs in reverse order and predict the next sequence of outputs of same time duration in future.}
	\label{fig:Iterative_Approach}
\end{figure}

To generate a long time-horizon dynamical prediction beyond the multi-timestep prediction by a single stacked deep LSTM neural network (shown as ``Deep LSTM'' in Figure \ref{fig:Iterative_Approach}), we used an iterative approach as described here. We made copies of the trained single stacked LSTM network and connected them in the feedforward manner in a sequence. We concatenated the sequence of predicted output from the previous stacked LSTM network with an equivalent length sequence of new inputs to the system and fed them in the reverse sequence order to the next stacked LSTM network. Figure \ref{fig:Iterative_Approach} illustrates this iterative approach.

\subsection{Sequence to Sequence Mapping with Neural Networks}\label{SS_NN}

\begin{figure}[htp]
	\centering
	\includegraphics[scale=0.5]{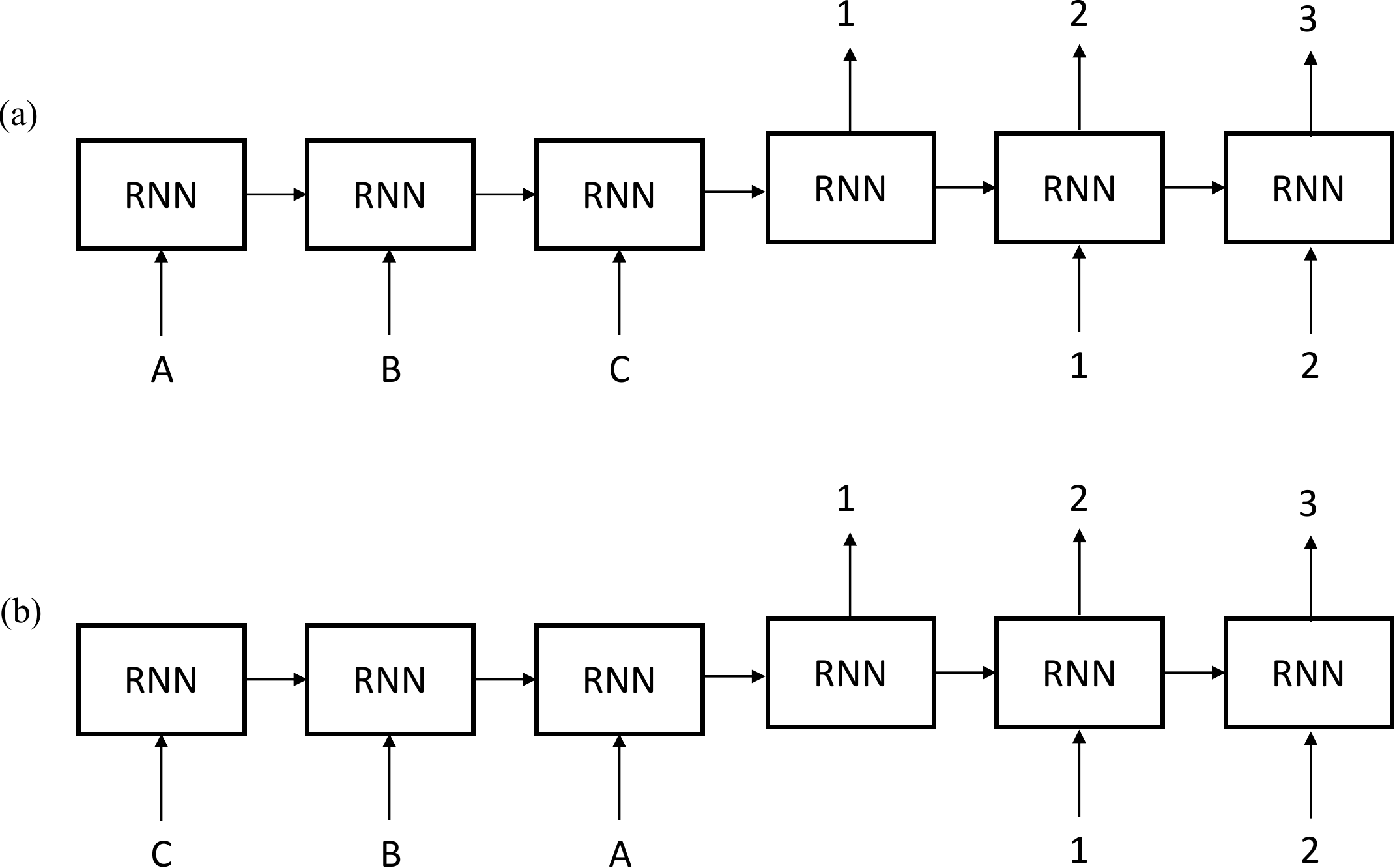}
	\caption{Forward and reversed sequence-to-sequence mapping approach for translating letters (inputs) to their numerical indices (outputs) in recurrent neural network (RNN). (a) shows the forward sequence-to-sequence mapping approach. The input is fed into the network in the same sequence as the desired output. The ``distance'' between all corresponding inputs and outputs is uniform. (b) shows the reversed sequence-to-sequence mapping approach. This approach introduces a temporal symmetry between input and output sequences while keeping the average ``distance'' between the corresponding inputs and outputs same as the forward approach. As shown in (b), $A\rightarrow1$ is the shortest "distance" to map, $B\rightarrow2$ the second, and $C\rightarrow3$ the furthest.}
	\label{fig:reverse_sequence}
\end{figure}

\noindent To make multi-timestep predictions of dynamical systems’ outputs using the deep LSTM neural network architecture described in the previous section (Section \ref{LSTM_NN}), we formulate the problem of mapping trajectories of the network inputs to the trajectories of the predicted outputs as a reverse order sequence-to-sequence mapping problem.  The central idea of the reverse order sequence-to-sequence mapping approach is to feed the inputs to the network in reverse order such that the network perceives the first input as the last and the last input as the first. Although this approach has been developed and applied in language translation applications \cite{sutskever2014sequence}, it has never been considered in the context of predicting dynamical systems behaviors from time-series data. Figure \ref{fig:reverse_sequence} illustrates the basic idea of the reverse order sequence-to-sequence mapping approach for translating letters (inputs) to their numerical indices (outputs).

As shown in Figure \ref{fig:reverse_sequence}, in the forward sequence-to-sequence mapping approach (Figure \ref{fig:reverse_sequence}(a)), i.e.,  $A,B,C\rightarrow1,2,3$, the distance between all mappings is same (i.e., 3 "units"). In the reverse sequence-to-sequence mapping approach (Figure \ref{fig:reverse_sequence}(b)), the network receives the input in a reverse order to map to the target output sequence, i.e., $C,B,A\rightarrow1,2,3$. As noted here, the average distance between the mappings remains the same for both approaches (i.e. 3 "units") but the reverse order approach introduces short and long-term symmetric temporal dependencies between inputs and outputs. These short and long-term symmetric temporal dependencies provide improved predictive performance over long temporal horizons \cite{sutskever2014sequence}.

\subsection{Synthetic Data}\label{syndata}
\noindent Hippocampal CA1 pyramidal neurons exhibit various multi-timescale firing patterns (from simple spiking to bursting) and play an essential role in shaping spatial and episodic memory \cite{ mckiernan2017ca1}. In the last two decades, several biophysiological models of the CA1 pyramidal (CA1Py) neurons ranging from single compartmental biophysiological and phenomenological models \cite{ golomb2006contribution,nowacki2011unified, ferguson2014simple} to detailed morphology-based multi-compartmental models \cite{ poirazi2003arithmetic, royeck2008role, katz2009synapse, bianchi2012mechanisms, marasco2012fast, kim2015dendritic, bezaire2016interneuronal} have been developed to understand the contributions of various ion-channels in diverse firing patterns (e.g., simple spiking to bursting) exhibited by the CA1Py neurons. 

\begin{figure}[htp]
	\centering
	\includegraphics[width=\textwidth]{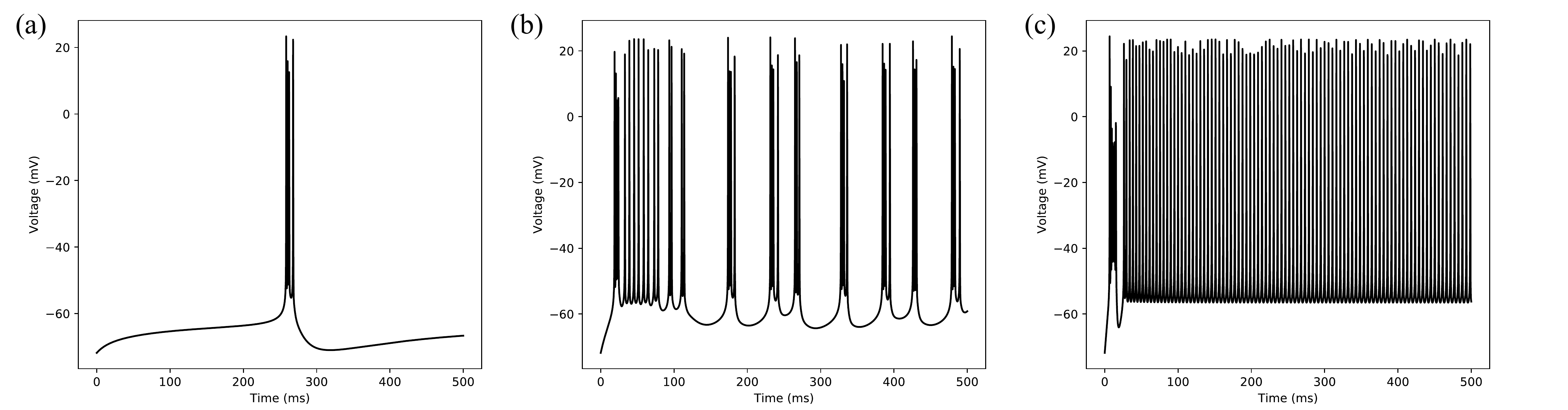}
	\caption{Diversity in the spiking patterns of hippocampal CA1 pyramidal neurons to applied currents. (a) Regular bursting in response to the external current of 0.23 nA. (b) Irregular bursting in response to the external current of 1.0 nA. (c) Plateau potentials followed by regular spiking in response to the external current of 3.0 nA.}
	\label{fig:spiking}
\end{figure}

In this paper, we use an experimentally validated 9-dimensional nonlinear model of CA1 pyramidal neuron in the Hodgkin-Huxley formalism given in \cite{golomb2006contribution} to generate the synthetic data for the network training and validation. The model exhibits several different bifurcations to the external stimulating current and has shown its capability in generating diverse firing patterns observed in electrophysiological recordings from CA1 pyramidal cells under various stimulating currents. Figure \ref{fig:spiking} shows three different firing patterns generated from this model based on the three different regimes of the applied input currents.     

To construct the synthetic training and validation dataset for the deep LSTM neural networks we designed in this paper with different predictive horizons, we simulated the Hodgkin-Huxley model of CA1 pyramidal neuron given in \cite{golomb2006contribution} (see Appendix \ref{HH_Model} for the details of the model) for 1000 ms duration for 2000 constant stimulating currents, sampled uniformly between $I = 0.0$ nA and $I=3.0$ nA. From these 2000 examples, we randomly and uniformly drew 50 samples (i.e., $10^4$ data points) of the desired predictive horizon as the input/output sequence data for training and validation. As described in Section \ref{training_NN}, we used 1/32 of these data points for validation, i.e., 96,875 data points for the training and 3,125 data points for the validation. Since our deep LSTM neural network takes an initial sequence of outputs of appropriate predictive horizon length (i.e., $N_{p} = 1,50,100,200$) as an input sequence to make the next time-horizon prediction of equivalent length of sequence, we assume that this initial output sequence data is available to the deep LSTM neural network throughout our simulations.

\subsection{Network Training}\label{training_NN}
\noindent We formulated the following optimization problem to train a set of network weights $\theta$: 
\begin{equation}
\theta^* = \argmin_{\theta} \mathcal{L}(\theta), 
\label{eq:optim_prob}
\end{equation}
where the loss function $\mathcal{L}(\theta)$ is given by
\begin{equation}
\mathcal{L}(\theta) = \frac{1}{N_P} \sum_{k=0}^{N_P}(\Vec{x}(k)-\hat{x}(k|\theta))^T(\Vec{x}(k)-\hat{x}(k|\theta)).
\label{eq:loss_function}
\end{equation}

\noindent Here $N_P$ represents the length of horizon over which the predictions are made, $\Vec{x}(k)$ is the known state vector at time step $k$, and $\hat{x}(k|\theta)$ is the neural network's prediction of the state vector at time $k$, given $\theta$.

To solve the optimization problem ~\eqref{eq:optim_prob}-\eqref{eq:loss_function}, we used the standard supervised backpropagation learning algorithm \cite{werbos1988generalization}, \cite{werbos1990backpropagation}, \cite{mozer1995focused} along with the Adaptive Moment Estimation (Adam) method \cite{kingma2014adam}. The Adam method is a first-order gradient-based optimization algorithm and uses lower-order moments of the gradients between layers to optimize a stochastic objective function.

Given the network parameter $\theta^{(i)}$ and the loss function $\mathcal{L}(\theta)$, where $i$ represents the algorithm's training iteration, the parameter update is given by \cite{kingma2014adam}
\begin{eqnarray}
m_\theta^{(i+1)} \leftarrow \beta_1 m_\theta^{(i)} + (1-\beta_1)\nabla_{\theta} \mathcal{L}^{(i)} \\
\nu_\theta^{(i+1)} \leftarrow \beta_2 m_\theta^{(i)} + (1-\beta_2)(\nabla_\theta \mathcal{L}^{(i)})^2 \\
\hat{m}_\theta = \frac{m_\theta^{(i+1)}}{1-(\beta_1)^{i+1}} \\
\hat{\nu}_\theta = \frac{\nu_\theta^{(i+1)}}{1-(\beta_2)^{(i+1)}} \\
\theta^{(i+1)} \leftarrow \theta^{(i)} - \eta \frac{\hat{m}_\theta}{\sqrt{\hat{\nu}_\theta}+\epsilon}
\end{eqnarray}
where $m_\theta$ is the first moment of the weights in a layer, $\nu_\theta$ is the second moment of the weights in a layer, $\eta$ is the learning rate, $\beta_1$ and $\beta_2$ are the exponential decay rates for the moment estimates, $\nabla$ is the differential gradient operator, and $\epsilon$ is a small scalar term to help numerical stability.  Throughout this work, we used $\beta_1 = 0.9$, $\beta_2 = 0.999$, and $\eta=0.001$ \cite{kingma2014adam}.

It should be noted that there is a tradeoff between the predictive time-horizon of deep LSTM neural network and the computational cost involved in training the network over the predictive horizon. As the predictive horizon increases, the computational cost of training the network over that horizon increases significantly for an equivalent number of examples. To keep the computational tractability in our simulations, all networks with long predictive horizons (i.e., $N_P = 50, 100, 200$) were trained for 200 epochs except the one-step predictive network which was trained for 1,000 epochs.

For all training sets throughout this paper, we used the validation to training data ratio as 1/32.  We set the minibatch size for training to 32. We performed all the training and computation in the TensorFlow computational framework on a discrete server running CentOS 7 with twin Nvidia GTX 1080Ti GPUs equipped with 11 Gb of VRAM.

\section{Simulation Results} \label{results}
\begin{figure}[htp]
	\centering
	\includegraphics[scale=1.1]{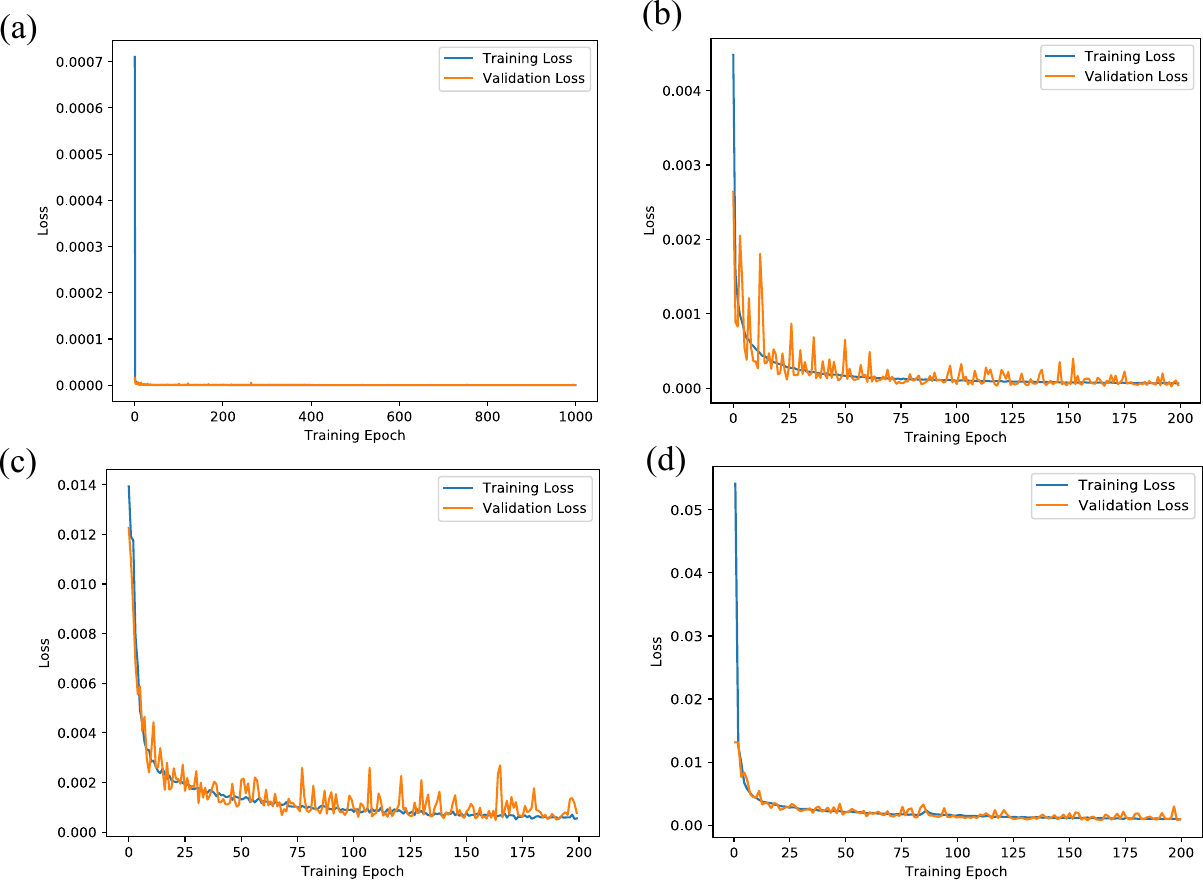}
	\caption{Training and validation loss for the deep LSTM neural network with multi-timestep predictive horizon. (a) 1 timestep predictive horizon. (b) 50 timesteps predictive horizon. (c) 100 timesteps predictive horizon. (d) 200 timesteps predictive horizon.}
	\label{fig:my_label}
\end{figure}

\noindent In this section, we present our simulation results on predicting the multi-timescale spiking dynamics exhibited by hippocampal CA1 pyramidal neurons over a long time-horizon using our developed deep LSTM neural network architecture described in Section \ref{Method}. We trained 4 LSTM networks for making one timestep prediction (equivalently, $0.1$ ms), 50 timesteps prediction (equivalently, $5$ ms), 100 timesteps prediction (equivalently, $10$ ms), and 200 timesteps prediction (equivalently, $20$ ms). Figure \ref{fig:my_label} shows the training and validation loss for these 4 LSTM networks.  

Using the iterative approach described in Section \ref{LSTM_NN}, we simulated the LSTM networks over $500$ ms of time duration under different initial conditions and stimulating input currents between three different regimes of dynamical responses (``Regular Spiking’’ ($I\in[2.3,3.0]$ nA), ``Irregular Bursting’’ ($I\in[0.79,2.3)$ nA), and ``Regular Bursting’’ ($I\in[0.24,0.79)$) nA) and compared the predicted state trajectories with the Hodgkin-Huxley model.

\subsection{Regular Spiking}\label{RS}
\noindent In this section, we demonstrate the efficacy of our trained deep LSTM neural network over the range of external current between $2.3$ nA and $3$ nA in predicting the regular spiking dynamics shown by the biophysiological Hodgkin-Huxley model of CA1 pyramidal neuron in response to the external current $I \geq 2.3$ nA. For clarity, we here show our results only for the membrane potential traces. We provide the complete set of simulation results on the LSTM network performance in predicting the dynamics of all the $9$ states of the Hodgkin-Huxley model in Appendix \ref{Appendix_RS} (see Figures \ref{fig:SS_Full-State_I=3,0}, \ref{fig:50S_Full-State_I=3,0}, \ref{fig:100S_Full-State_I=3,0}, \ref{fig:200S_Full-State_I=3,0}, and \ref{fig:RMSEvsTimeHighFull}).

\begin{figure}[htp]
	\centering
	\includegraphics[scale=0.7]{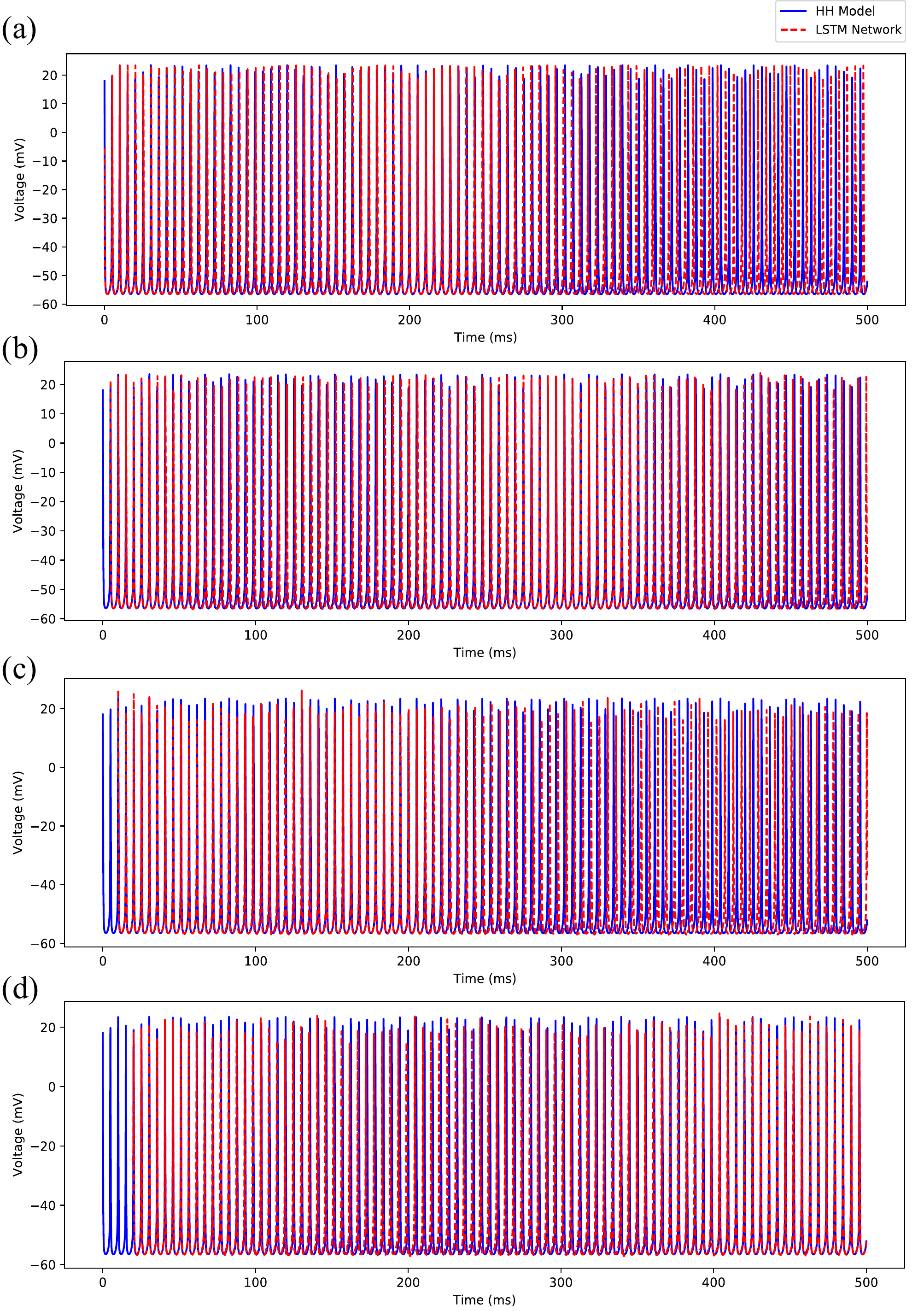}
	\caption{Comparison of predicted membrane potential traces by the deep LSTM neural network (``LSTM Network'') to the regular spiking pattern exhibited by the Hodgkin-Huxley model (``HH Model'') in response to the external stimulating current $I = 3.0$ nA. (a) Prediction using 1 timestep predictive LSTM network ($N_{p} = 1$). (b) Prediction using 50 timesteps predictive LSTM network ($N_{p} = 50$). (c) Prediction using 100 timesteps predictive LSTM network ($N_{p} = 100$). (d) Prediction using 200 timesteps predictive LSTM network ($N_{p} = 200$).}
	\label{fig:Comparison_I=3}
\end{figure}

Figure \ref{fig:Comparison_I=3} shows a comparison of the membrane potential traces simulated using the Hodgkin-Huxley model and the 4 different predictive horizons of the LSTM network (i.e., 1 timestep, 50 timesteps, 100 timesteps, and 200 timesteps, which we represent as $N_{p} = 1,50,100,200$) for the external stimulating current $I = 3.0$ nA. Note that all the simulations are performed using the same initial condition as provided in Appendix \ref{HH_Model}. Since our LSTM network uses the initial sequence of outputs of appropriate predictive horizon (i.e., $N_{p} = 1,50,100,200$) from the Hodgkin-Huxley model to make future time predictions,  the LSTM network predictions (shown by dashed red line) start after $0.1$ ms, $5$ ms, $10$ ms, and $20$ ms in Figure \ref{fig:Comparison_I=3}a, Figure \ref{fig:Comparison_I=3}b, Figure \ref{fig:Comparison_I=3}c, and Figure \ref{fig:Comparison_I=3}d respectively. 

\begin{figure}[ht]
	\centering
	\includegraphics[scale=0.7]{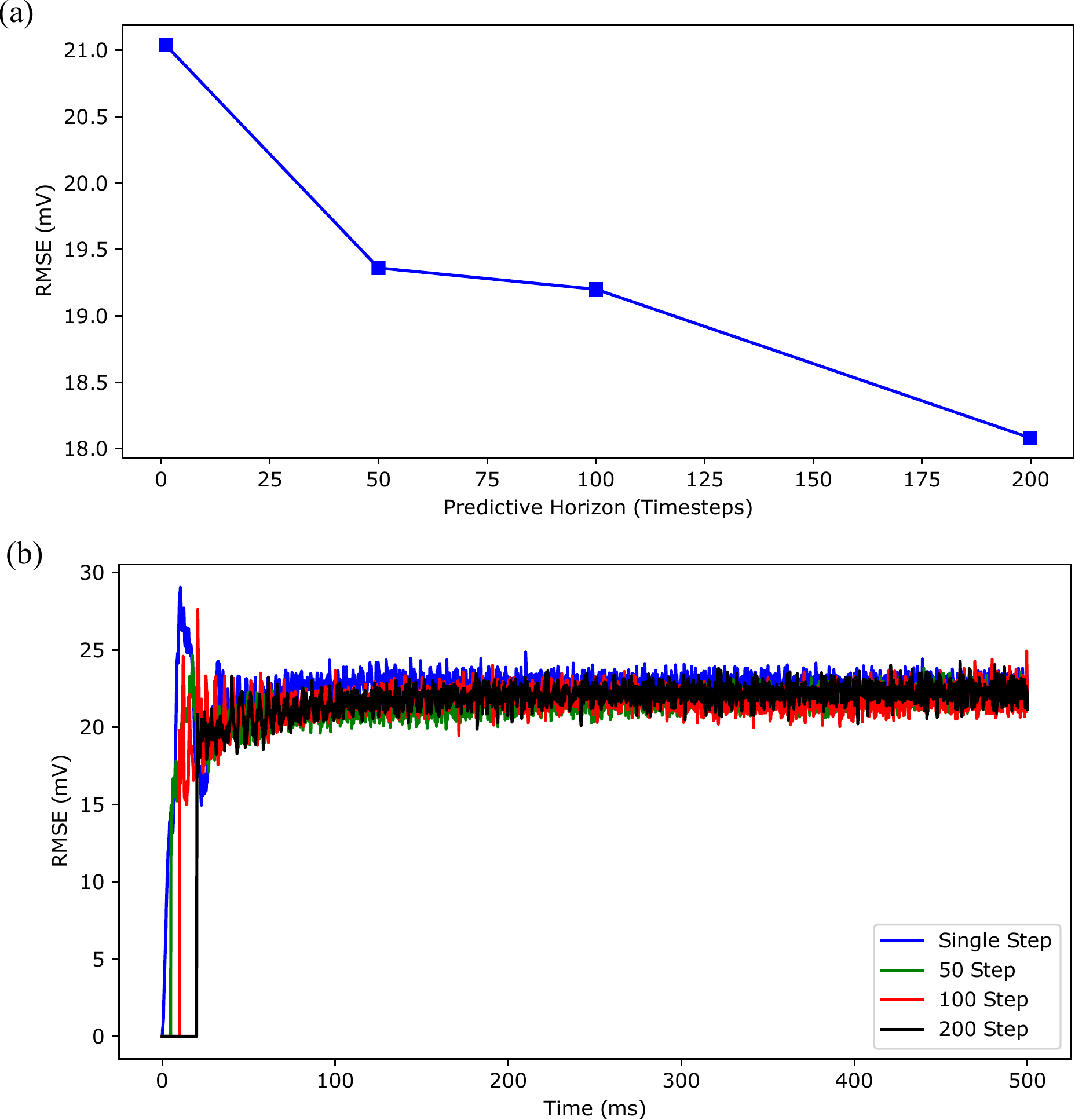}
	\caption{The effect of the length of predictive horizon of the deep LSTM neural network on the accuracy of regular spiking patterns prediction. (a) shows the time-averaged root mean squared error (RMSE) versus predictive horizon of the LSTM network ($N_{p} = 1, 50, 100, 200$) for the simulation results shown in Figure \ref{fig:Comparison_I=3}. (b) shows the RMSE versus simulation time for 5000 independent realizations, drawn from the predicted membrane potential trajectories of 50 randomly selected stimulating currents from a Uniform distribution $\mathcal{U}(2.3,3.0)$ and 100 random initial conditions for each stimulating current.}
	\label{fig:RMSE_RS_MP}
\end{figure}

As shown in Figure \ref{fig:Comparison_I=3}, the iterative prediction of the membrane potential traces by the LSTM network didn’t differ significantly over a short time horizon (up to 200 ms) for $N_{p} = 1,50,100,200$, but it significantly improved afterward with the increased predictive horizon of the LSTM network (i.e., $N_{p} = 1$ to $N_{p} = 200$). In particular, the LSTM network performance significantly improved in predicting the timing of the occurrence of spikes, but the magnitude of the membrane potential traces during spikes degraded as we increased $N_{p}$ from 1 to 200. For clarity, we also computed the time-averaged root mean squared error (RMSE) of the membrane potential traces between the Hodgkin-Huxley model and the LSTM network for $N_{p} = 1,50,100,200$ over 500 ms of simulation time. Figure \ref{fig:RMSE_RS_MP}(a) shows that the time-averaged RMSE decreased consistently with the increased predictive horizon of the LSTM network. Overall, these results show that our LSTM network with a longer predictive horizon prefers to capture temporal correlations more accurately over the amplitude while LSTM network with a shorter predictive horizon prefers to capture the amplitude more accurately over the temporal correlations.  

To systematically evaluate whether the designed LSTM networks provide reasonable predictions of the membrane potential traces of the regular spiking dynamics across the range of external input currents between $2.3$ nA and $3.0$ nA, we performed simulations for 50 random stimulating currents drawn from a Uniform distribution $\mathcal{U}(2.3,3.0)$. For each stimulating current, we chose 100 initial conditions drawn randomly from the maximum and minimum range of the Hodgkin-Huxley state variables (Note that the network was not trained over this wide range of initial conditions). Figure \ref{fig:RMSE_RS_MP}(b) shows the LSTM network performance, represented in terms of the root mean squared error vs time over $5000$ realizations, for $N_{p} = 1, 50, 100, 200$. As shown in this figure, the root mean squared error decreased with the increased predictive horizon of the LSTM network for all time, which is consistent with the result shown in Figure \ref{fig:RMSE_RS_MP}(a).  

In conclusion, these results suggest that our deep LSTM neural network with a longer predictive horizon feature can predict the regular (periodic) spiking patterns exhibited by hippocampal CA1 pyramidal neurons with high accuracy over a long-time horizon. 

\subsection{Irregular Bursting}\label{IB}
\noindent In this section, we demonstrate the efficacy of our trained deep LSTM neural network over the range of external current between $0.79$ nA and $2.3$ nA in predicting the irregular bursting dynamics shown by the biophysiological Hodgkin-Huxley model of CA1 pyramidal neuron in response to the external current $I \in [0.79,2.3)$ nA. For clarity, we here show our results only for the membrane potential traces. We provide the complete set of simulation results on the LSTM network performance in predicting the dynamics of all the $9$ states of the Hodgkin-Huxley model in Appendix \ref{Appendix_IBS} (see Figures \ref{fig:SS_Full-State_I=1.5}, \ref{fig:50S_Full-State_I=1.5}, \ref{fig:100S_Full-State_I=1.5}, \ref{fig:200S_Full-State_I=1.5}, and \ref{fig:RMSEvsTimeMediumFull}).

\begin{figure}[htp]
	\centering
	\includegraphics[scale=0.7]{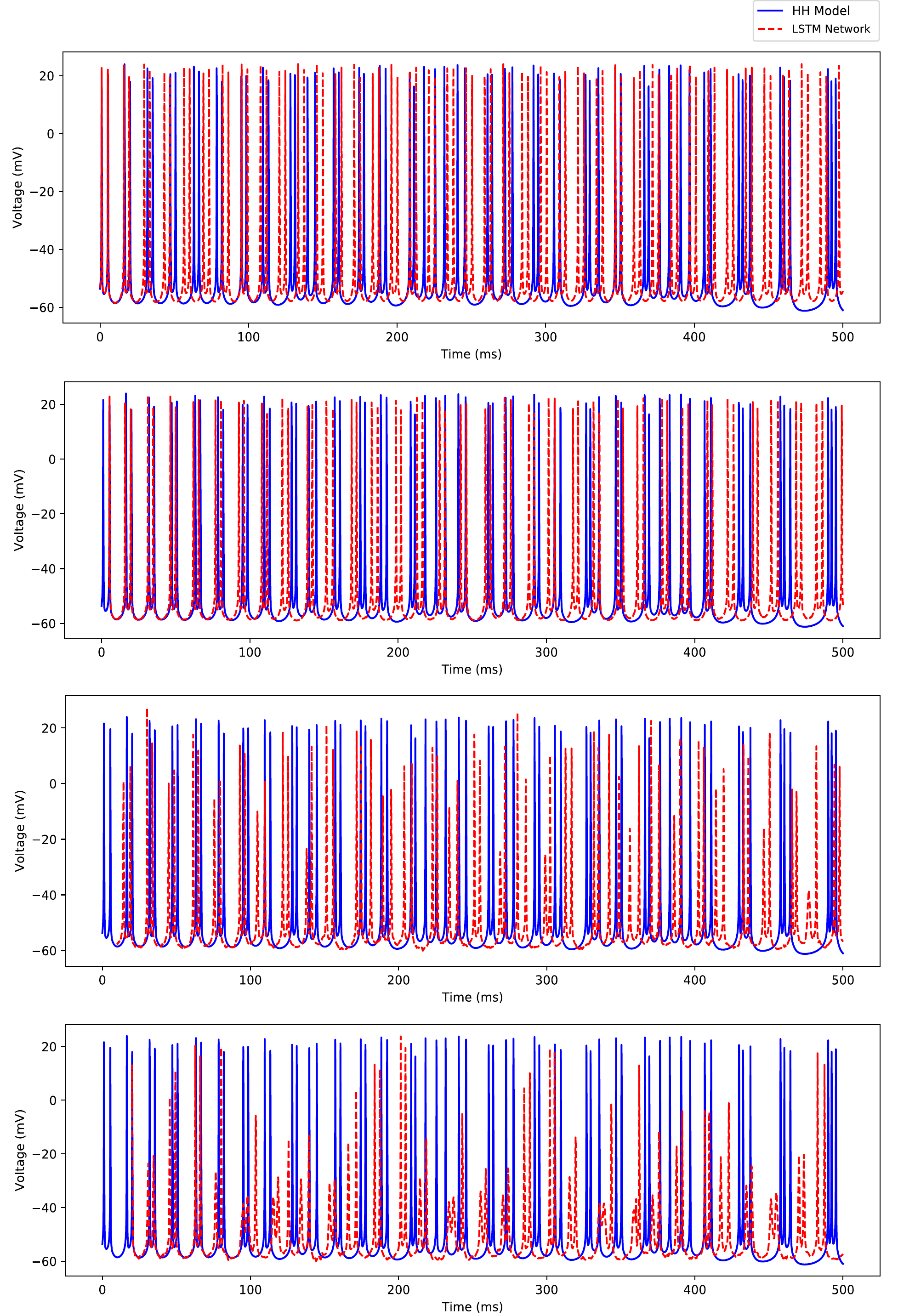}
	\caption{Comparison of predicted membrane potential traces by the deep LSTM neural network (``LSTM Network'') to the irregular bursting spiking patterns exhibited by the Hodgkin-Huxley model (``HH Model'') in response to the external stimulating current $I = 1.5$ nA. (a) Prediction using 1 timestep predictive LSTM network ($N_{p} = 1$). (b) Prediction using 50 timesteps predictive LSTM network ($N_{p} = 50$). (c) Prediction using 100 timesteps predictive LSTM network ($N_{p} = 100$). (d) Prediction using 200 timesteps predictive LSTM network ($N_{p} = 200$).}
	\label{fig:Comparison_I=1.5}
\end{figure}

Figure \ref{fig:Comparison_I=1.5} shows a comparison of the membrane potential traces simulated using the Hodgkin-Huxley model and the 4 different predictive horizons of the LSTM network (i.e., $N_{p} = 1,50,100,200$) for the external stimulating current $I = 1.5$ nA. Note that all the simulations are performed using the initial condition used for $I=3.0$ nA  in Figure \ref{fig:Comparison_I=3}. Since our LSTM network uses the initial sequence of outputs of appropriate prediction horizon (i.e., $N_{p} = 1, 50, 100, 200$) from the Hodgkin-Huxley model to make future time predictions, the LSTM network predictions (shown by dashed red line) start after $0.1$ ms, $5$ ms, $10$ ms, and $20$ ms in Figure \ref{fig:Comparison_I=1.5}a, Figure \ref{fig:Comparison_I=1.5}b, Figure \ref{fig:Comparison_I=1.5}c, and Figure \ref{fig:Comparison_I=1.5}d respectively. 

\begin{figure}[h]
	\centering
	\includegraphics[scale=0.7]{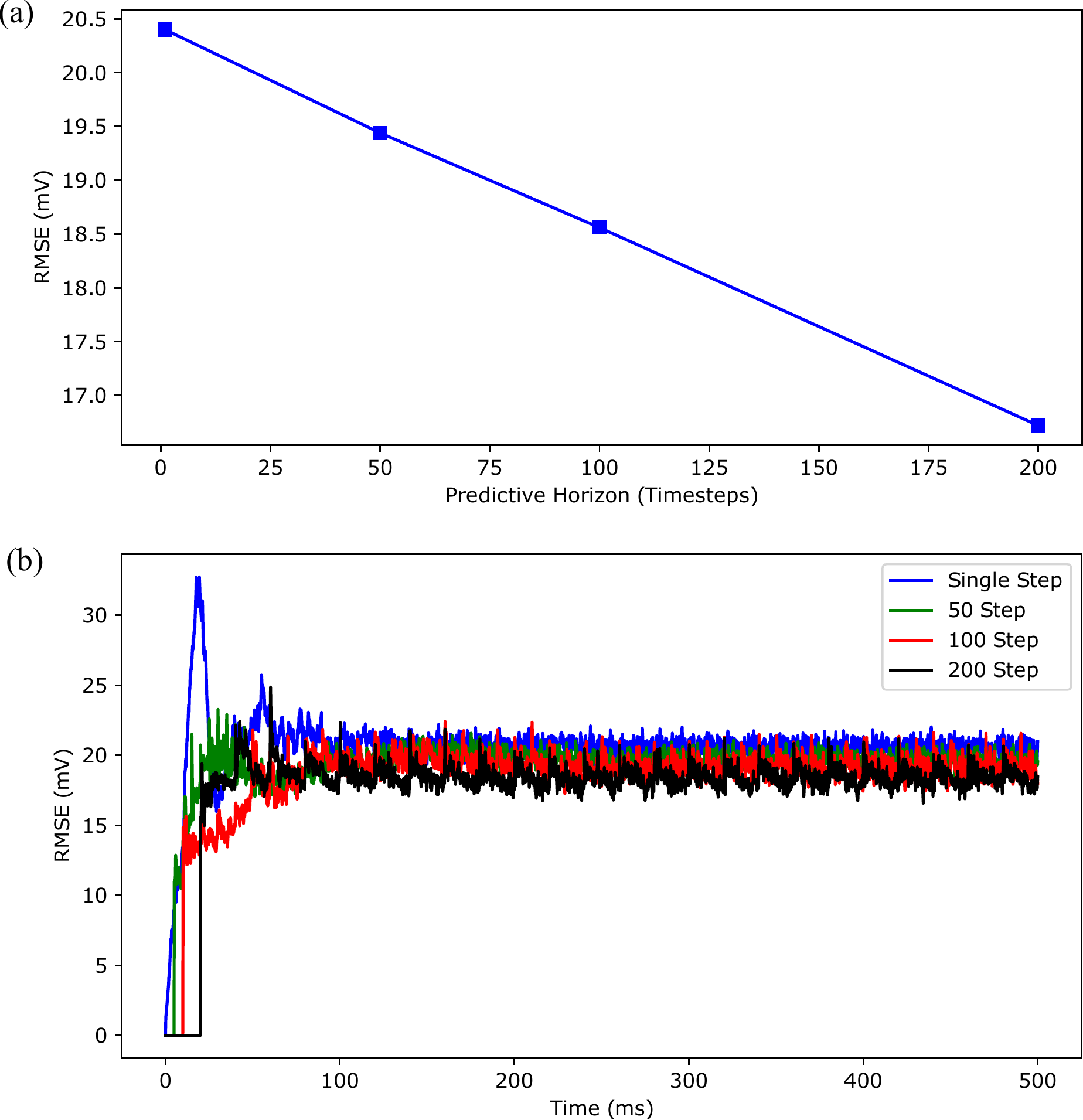}
	\caption{The effect of the prediction horizon of the deep LSTM neural network on the accuracy of irregular bursting dynamics prediction. (a) shows the time-averaged root mean squared error (RMSE) versus predictive horizon of the LSTM network ($N_{p} = 1, 50, 100, 200$) for the simulation results shown in Figure \ref{fig:Comparison_I=1.5}. (b) shows the RMSE versus simulation time for 5000 independent realizations, drawn from the predicted membrane potential trajectories of 50 randomly selected stimulating currents from a Uniform distribution $\mathcal{U}(0.79,2.3)$ and 100 random initial conditions for each stimulating current.}
	\label{fig:RMSE_IBS_MP}
\end{figure}

As shown in Figure \ref{fig:Comparison_I=1.5}, the LSTM performance significantly improved in predicting the timing of the occurrence of spikes up to 100 ms with the increased predictive horizon of the LSTM network from $N_{p} = 1$ to $N_{p} = 200$, but the performance degraded in capturing the magnitude of the membrane potentials during spiking with an increased value of $N_{p}$. Although the time-averaged root mean squared error of the membrane potential traces between the Hodgkin-Huxley model and the LSTM network for $N_{p} = 1, 50, 100, 200$ showed an improved performance with the increased value of $N_{p}$ (see Figure  \ref{fig:RMSE_IBS_MP}(a)), none of the LSTM networks showed a reasonable prediction of the timing of the occurrence of spikes in this regime beyond 100 ms of the time-horizon.

To systematically evaluate whether the designed LSTM networks provide reasonable predictions of the membrane potential traces of the regular spiking dynamics across the range of external input currents between $0.79$ nA and $2.3$ nA, we performed simulations for 50 random stimulating currents drawn from a Uniform distribution $\mathcal{U}(0.79,2.3)$. For each stimulating current, we chose 100 initial conditions drawn randomly from the maximum and minimum range of the Hodgkin-Huxley state variables (Note that the network was not trained over this wide range of initial conditions). Figure \ref{fig:RMSE_IBS_MP}(b) shows the LSTM network performance, represented in terms of the root mean squared error vs time over $5000$ realizations, for $N_{p} = 1, 50, 100, 200$. As shown in this figure, the root mean squared error decreased with the increased predictive horizon of the LSTM network for all time, which is consistent with the result shown in Figure \ref{fig:RMSE_IBS_MP}(a).  

In conclusion, these results suggest that our deep LSTM neural network with a longer predictive horizon feature can predict the irregular bursting patterns exhibited by hippocampal CA1 pyramidal neurons with high accuracy over only a short-time horizon.

\subsection{Regular Bursting}
\noindent  In this section, we demonstrate the efficacy of our trained deep LSTM neural network over the range of external current between $0.24$ nA and $0.79$ nA in predicting the regular bursting dynamics shown by the biophysiological Hodgkin-Huxley model of CA1 pyramidal neuron in response to the external current $I \in [0.24,0.79)$ nA. For clarity, we here show our results only for the membrane potential traces. We provide the complete set of simulation results on the LSTM network performance in predicting the dynamics of all the $9$ states of the Hodgkin-Huxley model in Appendix \ref{Appendix_RBS} (see Figures \ref{fig:SS_Full-State_I=0.5}, \ref{fig:50S_Full-State_I=0.5}, \ref{fig:100S_Full-State_I=0.5}, \ref{fig:200S_Full-State_I=0.5}, and \ref{fig:RMSEvsTimeLowFull}).

\begin{figure}
	\centering
	\includegraphics[scale=0.7]{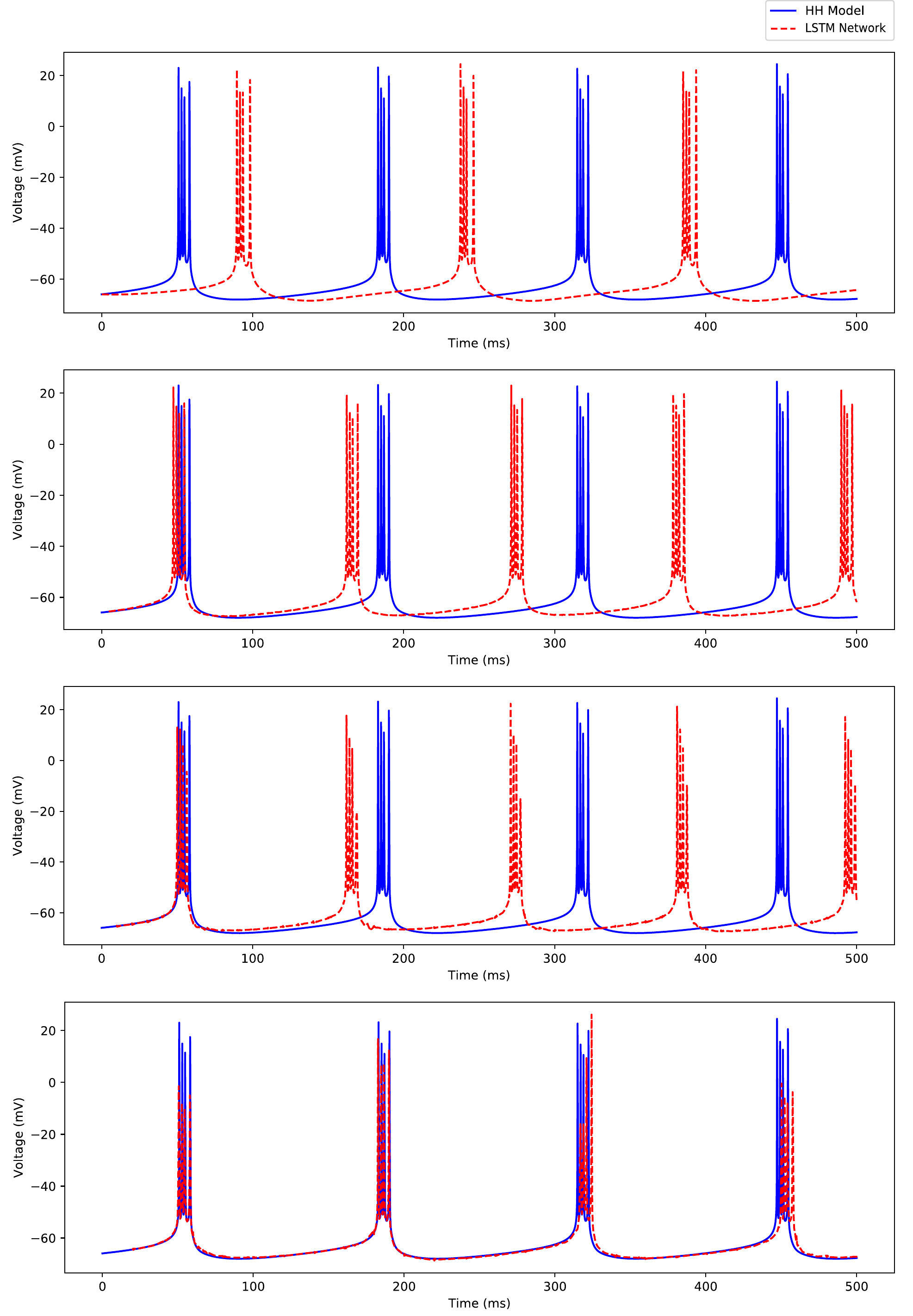}
	\caption{Comparison of predicted membrane potential traces by the LSTM network (``NN Prediction'') to the irregular bursting spiking patterns exhibited by the Hodgkin-Huxley model (``HH Model'') in response to the external stimulating current $I = 0.5$ nA. (a) Prediction using 1 timestep predictive LSTM network ($N_{p} = 1$). (b) Prediction using 50 timesteps predictive LSTM network ($N_{p} = 50$). (c) Prediction using 100 timesteps predictive LSTM network ($N_{p} = 100$). (d) Prediction using 200 timesteps predictive LSTM network ($N_{p} = 200$).}
	\label{fig:Comparison_I=0.5}
\end{figure}

Figure \ref{fig:Comparison_I=0.5} shows a comparison of the membrane potential traces simulated using the Hodgkin-Huxley model and the 4 different predictive horizons of the LSTM network (i.e., $N_{p} = 1,50,100,200$) for the external stimulating current $I = 0.5$ nA. Note that all the simulations are performed using the initial condition used for $I=3.0$ nA  in Figure \ref{fig:Comparison_I=3}. Since our LSTM network uses the initial sequence of outputs of appropriate prediction horizon (i.e., $N_{p} = 1, 50, 100, 200$) from the Hodgkin-Huxley model to make future time predictions, the LSTM network predictions (shown by dashed red line) start after $0.1$ ms, $5$ ms, $10$ ms, and $20$ ms in Figure \ref{fig:Comparison_I=0.5}a, Figure \ref{fig:Comparison_I=0.5}b, Figure \ref{fig:Comparison_I=0.5}c, and Figure \ref{fig:Comparison_I=0.5}d respectively.  

By analyzing the results shown in Figure \ref{fig:Comparison_I=0.5}, we found that the LSTM network performance in predicting the timing of spikes during bursts as well as tracking the subthreshold potential improved significantly from $N_{p}=1$ to $N_{p} = 200$, but the performance substantially degraded in capturing the magnitude of the membrane potentials during spiking. In conclusion, the 200 timesteps prediction horizon based LSTM network (see Figure \ref{fig:Comparison_I=0.5}(d)) predicts the temporal dynamics with reasonable accuracy over the first 300 ms of prediction. 

\begin{figure}[h]
	\centering
	\includegraphics[scale=0.7]{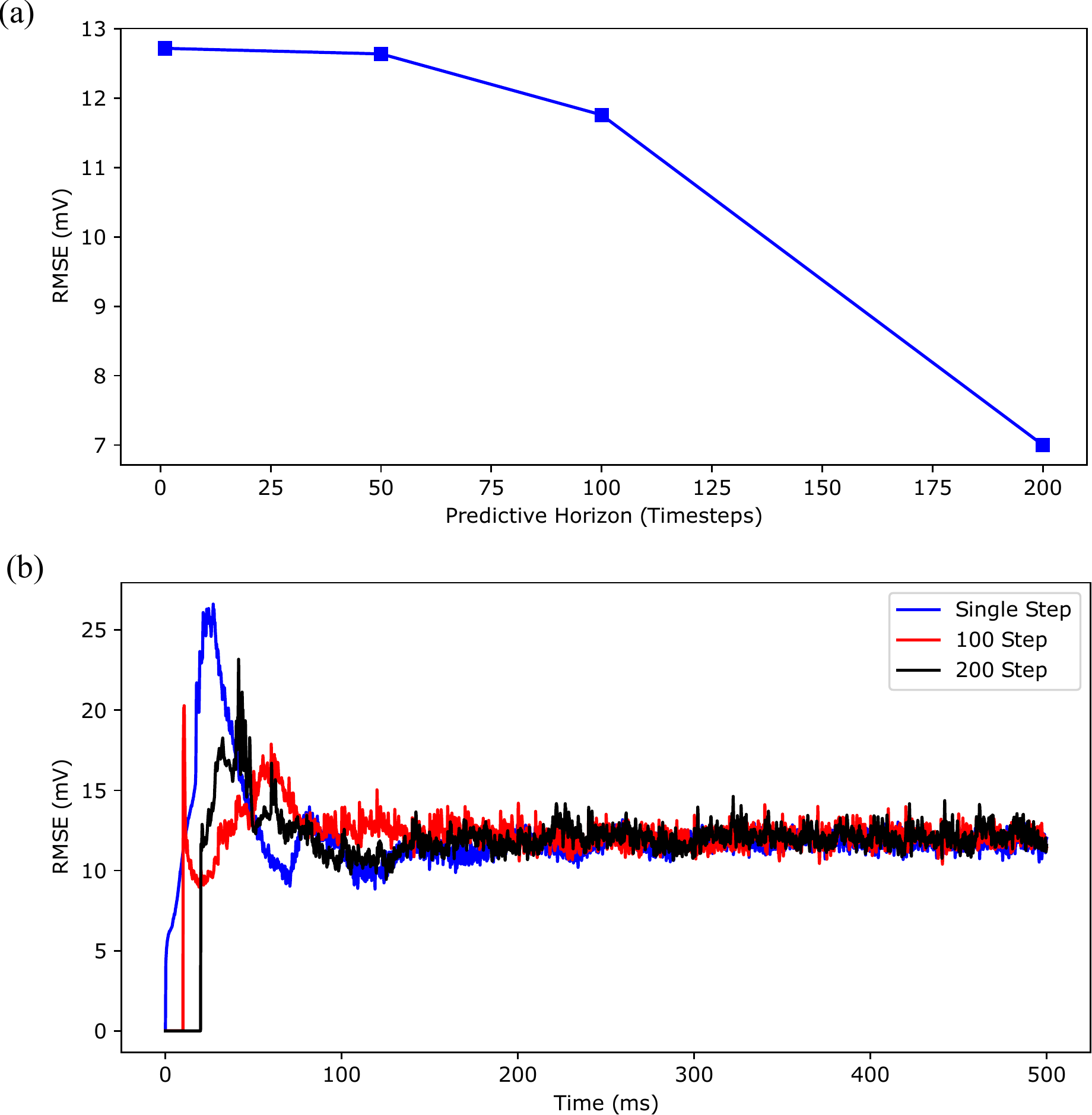}
	\caption{The effect of the prediction horizon of the multi-timestep LSTM network on the accuracy of regular bursting dynamics prediction. (a) shows the time-averaged root mean squared error (RMSE) versus predictive horizon of the LSTM network ($N_{p} = 1, 50, 100, 200$) for the simulation results shown in Figure \ref{fig:Comparison_I=0.5}. (b) shows the RMSE versus simulation time for 5000 independent realizations, drawn from the predicted membrane potential trajectories of 50 randomly selected stimulating currents from a Uniform distribution $\mathcal{U}(0.24,0.79)$ and 100 random initial conditions for each stimulating current.}
	\label{fig:RMSE_RBS_MP}
\end{figure}

Figure \ref{fig:RMSE_RBS_MP}(a) shows the time-averaged root mean squared error of the membrane potential traces between the Hodgkin-Huxley model and the LSTM network for $N_{p} = 1, 50, 100, 200$. As noted in this figure, the root mean squared error decreased substantially between 100 timesteps and 200 timesteps prediction horizon compared to the regimes of regular spiking (Figure \ref{fig:RMSE_RS_MP}(a)) and irregular bursting (Figure \ref{fig:RMSE_IBS_MP}(a)), which indicates that a longer predictive horizon based LSTM network is necessary to capture the two different timescales (i.e., short intraburst spiking intervals and long interburst subthreshold intervals) presented in these dynamics.      

Figure \ref{fig:RMSE_RBS_MP}(b) shows the LSTM networks performances, represented in terms of the root mean squared error vs time over $5000$ realizations, for $N_{p} =$ 1, 100, and 200 timestep prediction horizon LSTM network. As shown in this figure, the root mean squared error decreased with the increased predictive horizon of the LSTM network for all time, which is consistent with the result shown in Figure \ref{fig:RMSE_RBS_MP}(a). Note that we have excluded the simulation result for $N_{p} = 50$ as we found out in our detailed analysis that the trained LSTM network for $N_{p} = 50$ led to instability in predicting spiking responses for some of the initial condition values in this regime. The reason for this may be that the network may not have seen these initial conditions during the training.

\section{Discussion}\label{discussion}
\noindent In this paper, we developed and presented a new data-driven long short-term memory (LSTM) based neural network (NN) architecture to predict the dynamical spiking patterns of single neurons. Compared to other LSTM-based NN architectures for forecasting dynamical systems behavior reported in the literature, our architecture incorporated a single dense feedforward output layer with an activation function and a reverse-order sequence-to-sequence mapping approach into traditional LSTM based neural networks to enable truly multi-timestep stable predictions of the dynamics over a long time-horizon. We demonstrated the efficacy of our architecture in predicting the multi-time scale dynamics of hippocampal CA1 pyramidal neurons and compared the predictions from our model with the ground truth synthetic data obtained from an experimentally validated biophysiological model of CA1 pyramidal neuron in the Hodgkin-Huxley formalism. Through simulations, we showed that (1) the presented architecture can learn multi-timescale dynamics, and (2) the predictive accuracy of the network increases with the increase in the predictive horizon of the LSTM network. 

Our results for irregular bursting regime showed the limitation of the designed deep LSTM neural network architecture in making an accurate prediction of the timing of the occurrence of spikes over a long-time horizon compared to regular spiking and regular bursting regimes. A possible reason for this may be the architecture itself or the dataset used for training these networks, which requires further investigations by training the networks on the dataset explicitly generated from this regime.

In all dynamical regimes, our results showed a degraded performance of the deep LSTM neural network in predicting the amplitude of membrane potentials during the timing of the occurrence of spikes with the increased predictive horizon of the LSTM network. This issue may be related to the equally weighted norm-2 loss function used for training the networks. A further investigation is required by considering different loss functions, such as norm-1 or weighted norm-2, which we consider as our future work.

Although the data-driven approach developed in this paper showed the ability of the designed LSTM-based neural network in learning multi-timescale dynamics, we note that the network struggles to accurately capture the dynamics of some state variables where the magnitude of the state variable is comparable to the numerical precision of our simulations. This can particularly be seen in \ref{fig:50S_Full-State_I=3,0}, \ref{fig:100S_Full-State_I=3,0}, \ref{fig:200S_Full-State_I=3,0}, and \ref{fig:200S_Full-State_I=0.5}, where the network is not able to reconstruct the dynamics of the state variable $q_{sAHP}$ with a reasonable accuracy.  One possible way to alleviate this issue may be to increase the tolerance of the numerical errors in simulations which may increase the overall computational cost during training.

In conclusion, our results showed that a longer predictive horizon-based LSTM network can provide a more accurate prediction of multi-time scale dynamics, but at the expense of extensive offline training cost.

\bibliographystyle{unsrt}  
\bibliography{References}  

%
%
%
%

\appendix
\section{Hodgkin-Huxley Model of CA1 Pyramidal Neuron Dynamics}\label{HH_Model}

\noindent We used the following Hodgkin-Huxley model of CA1 pyramidal neuron from \cite{golomb2006contribution} to demonstrate the efficacy of our data-driven modeling approach presented in this paper:    
\begin{equation}
C\frac{dV}{dt} = -g_L(V-V_L) - I_{Na} - I_{NaP} - I_{Kdr} - I_A - I_M - I_{Ca} - I_C - I_{sAHP} + I_{app}, 
\end{equation}

\noindent where the ionic currents $I_{Na}$, $I_{NaP}$, $I_{Kdr}$, $I_{A}$, $I_{M}$, $I_{sAHP}$, $I_{C}$, and $I_{Ca}$ are given by 

\begin{subequations}
	\begin{equation}
	I_{Na} = g_{Na} m_{\infty}^3(V) h_{Na} (V-V_{Na}),
	\end{equation}	
	\begin{equation}
	I_{NaP} = g_{NaP} p_{\infty}(V) (V-V_{Na}),
	\end{equation}
	\begin{equation}
	I_{Kdr} = g_{Kdr} n_{Kdr}^4 (V-V_K),
	\end{equation}
	\begin{equation}
	I_A = g_A a_{\infty}^3(V) b_{Kdr} (V-V_K),
	\end{equation}
	\begin{equation}
	I_M = g_M z_M (V-V_K),
	\end{equation}
	\begin{equation}
	I_{Ca} = g_{Ca} r_{Ca}^2 (V-V_{Ca}),
	\end{equation}	
	\begin{equation}
	I_C = g_C d_{\infty} ([Ca^{2+}]_i) c_C (V-V_K),
	\end{equation}
	\begin{equation}
	I_{sAHP} = g_{sAHP} q_{sAHP} (V-V_K).
	\end{equation}
\end{subequations}

\noindent Here, $V$ is the membrane potential in mV, $C$ is the membrane capacitance, $V_{L}$ is the reversal potential of the leak current, $g_{L}$ is the conductance of the leak current, and $I_{app}$ is the externally applied stimulating current. The ionic currents $I_{Na}$, $I_{NaP}$, $I_{Kdr}$, $I_{A}$, $I_{M}$, $I_{sAHP}$, $I_{C}$, and $I_{Ca}$ represent the transient sodium current, persistent sodium current, delayed rectifier potassium current, A-type potassium current, muscarinic-sensitive potassium current, slow calcium-activated potassium current, fast calcium-activated potassium current, and high threshold calcium current respectively. $g_{i}$, $i\in\{Na,NaP,Kdr,A,M,Ca,C,sAHP\}$ represents the conductance of the ion channel $i$. $V_{i}$, $i\in \{Na, K, Ca\}$ is the reversal potential of the ion channel $i$. 

The dynamics of the transient activation/deactivation variables of the ionic and calcium currents, i.e., $h_{Na}$, $n_{Kdr}$, $b_{Kdr}$, $z_{M}$, $r_{Ca}$, $c_{C}$, $q_{sAHP}$, and $[Ca^{2+}]_{i}$, are given by:

\begin{subequations}
	\begin{equation}
	\frac{dh_{Na}}{dt} = \phi \frac{h_{\infty}(V) - h_{Na}}{\tau_{h_{Na}}(V)},
	\end{equation}
	\begin{equation}
	\frac{dn_{Kdr}}{dt} = \phi \frac{n_{\infty}(V) - n_{Kdr}}{\tau_{n_{Kdr}}(V)}, 
	\end{equation}
	\begin{equation}
	\frac{db_{Kdr}}{dt} = \frac{b_{\infty}(V)-b_{Kdr}}{\tau_{b_{Kdr}}}, 
	\end{equation}	 
	\begin{equation}
	\frac{dz_M}{dt} = \frac{z_{\infty}(V)-z_M}{\tau_z},  
	\end{equation}
	\begin{equation}
	\frac{dr_{Ca}}{dt} = \frac{r_{\infty}(V) - r_{Ca}}{\tau_{r_{Ca}}},  
	\end{equation}
	\begin{equation}
	\frac{dc_C}{dt} = \frac{c_{\infty}(V) - c_C}{\tau_{c_C}},  
	\end{equation}
	\begin{equation}
	\frac{dq_{sAHP}}{dt} = \frac{q_{\infty}(V) - q_{sAHP}}{\tau_{q_{sAHP}}},  
	\end{equation}
	\begin{equation}
	\frac{d[Ca^{2+}]_i}{dt} = -\nu I_{Ca} - \frac{[Ca^{2+}]_i}{\tau_{Ca}}.  
	\end{equation}
\end{subequations}

\noindent Here, $m_{\infty}(V)$, $h_{\infty}(V)$, $n_{\infty}(V)$, $p_{\infty}(V)$, $a_{\infty}(V)$, $b_{\infty}(V)$, $z_{\infty}(V)$, $r_{\infty}(V)$, $c_{\infty}(V)$, $q_{\infty}([Ca^{2+}]_i)$, and $d_{\infty}([Ca^{2+}]_i)$ are the steady-state activation/deactivation functions. $\phi$ is a scaling parameter. $\tau_{h_{Na}}(V)$, $\tau_{n_{Kdr}}(V)$, $\tau_{b_{Kdr}}(V)$, $\tau_{r_{Ca}}$, $\tau_{c_{C}}$, and $\tau_{q_{sAHP}}$ are the time constants. The steady-state activation/deactivation functions are given by:

\begin{subequations}
	\begin{equation}
	m_{\infty}(V) = \frac{1}{1+e^{-(V-\theta_m)/\sigma_m}},
	\end{equation}
	\begin{equation}
	n_{\infty}(V) = \frac{1}{1+e^{-(V-\theta_n)/\sigma_n}},
	\end{equation}
	\begin{equation}
	h_{\infty}(V) = \frac{1}{1+e^{-(V-\theta_h)/\sigma_h}},
	\end{equation}
	\begin{equation}
	p_{\infty}(V) = \frac{1}{1+e^{-(V-\theta_p)/\sigma_p}},
	\end{equation}
	\begin{equation}
	b_{\infty}(V) = \frac{1}{1+e^{-(V-\theta_b)/\sigma_b}},
	\end{equation}
	\begin{equation}
	z_{\infty}(V) = \frac{1}{1+e^{-(V-\theta_z)/\sigma_z}},
	\end{equation}
	\begin{equation}
	a_{\infty}(V) = \frac{1}{1+e^{-(V-\theta_a)/\sigma_a}},
	\end{equation}
	\begin{equation}
	r_{\infty}(V) = \frac{1}{1+e^{-(V-\theta_r)/\sigma_r}},
	\end{equation}
	\begin{equation}
	c_{\infty}(V) = \frac{1}{1+e^{-(V-\theta_c)/\sigma_c}},
	\end{equation}
	\begin{equation}
	d_{\infty}([Ca^{2+}]_i) = \frac{1}{(1+a_c/[Ca^{2+}]_i)}, 
	\end{equation}
	\begin{equation}
	q_{\infty}([Ca^{2+}]_i) = \frac{1}{1+(a_q^4/[Ca^{2+}]_i^4)}.
	\end{equation}
\end{subequations}

\noindent Here, $a_{c}$, $a_{q}$, $\theta_{i}$, $\sigma_{i}$ for $i\in\{m,n,h,p,b,z,a,r,c\}$ are the model parameters. The voltage dependent time constants $\tau_{h_{Na}}(V)$ and $\tau_{n_{Kdr}}(V)$ are given by 

\begin{subequations}
	\begin{equation}
	\tau_{h_{Na}}(V) = 1 + \frac{7.5}{1+e^{-(V-\theta_{ht})/\sigma_{ht}}},
	\end{equation}
	\begin{equation}
	\tau_{n_{Kdr}}(V) = 1 + \frac{5}{1+e^{-(V-\theta_{nt})/\sigma_{nt}}}, 
	\end{equation}s
\end{subequations}

\noindent where $\theta_{ht}$, $\theta_{nt}$, $\sigma_{ht}$, and $\sigma_{nt}$ are model parameters. 

Throughout this paper, we used the following numerical values for the unknown model parameters \cite{golomb2006contribution}: 
$C = 1 \mu$F/cm$^2$, $g_L = 0.05$ mS/cm$^2$, $V_L = -70$ mV, $\nu=0.13$ cm$^2$/(ms $\times$ $\mu$A), $g_Na$ = 35 mS/cm$^2$, $V_Na$ = 55 mV, $g_{NaP}$ = 0.4 mS/cm$^2$, $g_{Kdr}$ = 6.0 mS/cm$^2$, $V_K$ = -90 mV, $g_A$ = 1.4 mS/cm$^2$, $g_M$ = 0.5 mS/cm$^2$, $g_{Ca} =$ 0.08 mS/cm$^2$, $g_C$ = 10 mS/cm$^2$, $V_{Ca}$ = 120 mV,and $g_{sAHP}$ = 5 mS/cm$^2$, $\theta_m$ = -30 mV, $\sigma_m$ = 9.5 mV, $\theta_h$ = -45 mV, $\sigma_h$ = -7 mV, $\theta_{ht}$ = -40.5 mV, $\sigma_{ht}$ = -6 mV, $\phi$ = 10, $\theta_P$ = -47 mV, $\sigma_P$ = 3 mV, $\theta_n$ = -35 mV, $\sigma_n$ = 10 mV, $\theta_{nt}$ = -27 mV, $\sigma_{nt}$ = -15 mV, $\theta_a$ = -50 mV, $\sigma_a$ = 20 mV, $\theta_b$ = -80 mV, $\sigma_b$ = -6 mV, $\theta_z$ = -39 mV, $\sigma_z$ = 5 mV, $\theta_r$ = -20 mV, $\sigma_r$ = 10 mV, $\tau_r$ = 1 ms, $\theta_c$ = -30 mV, $\sigma_c$ = 7 mV, $\theta_c$ = 2 ms, $a_c$ = 6, $\tau_q$ = 450 ms, and $a_q$ = 2.

Unless otherwise stated, we used the following initial conditions to simulate the Hodgkin-Huxley model for generating the synthetic data: $V_0$ = -71.81327mV, $h_{Na0} = 0.98786$, $n_{Kdr0} = 0.02457$, $b_{KA0} = 0.203517$, $u_{KM0} = 0.00141$, $r_{Ca0} = 0.005507$, $[Ca]_{i0} = 0.000787$, $c_{C0} = 0.002486$, $q_{Ca0} = 0.0$.

\section{Simulation Results on Full State Predictions of Hodgkin-Huxley Model}
\noindent In Section \ref{results}, we showed our simulation results only for the membrane potential traces. Here, we provide the simulation results for all the 9 states of the Hodgkin-Huxley model of CA1 pyramidal neuron (HHCA1Py) predicted by the deep LSTM neural network over a long-time horizon and show the comparison between these predictions and the simulated dynamics from HHCA1Py. 

\subsection{Regular Spiking} \label{Appendix_RS}
\noindent In this section, we show the simulation results on predicting the dynamics of all the 9 states of HHCA1Py over a long-time horizon using the deep LSTM neural network for the regular periodic spiking regime ($I\in[2.3,3.0]$ nA).  Figures \ref{fig:SS_Full-State_I=3,0}, \ref {fig:50S_Full-State_I=3,0}, \ref {fig:100S_Full-State_I=3,0}, and \ref {fig:200S_Full-State_I=3,0} show the comparison between the state's dynamics simulated using the Hodgkin-Huxley model and the deep LSTM neural network model developed for 1 timestep, 50 timesteps, 100 timesteps, and 200 timesteps (equivalently, $N_{p} = 1,50,100,200$) predictive horizon, respectively.

\begin{figure}[htp]
	\centering
	\includegraphics[scale = 0.5]{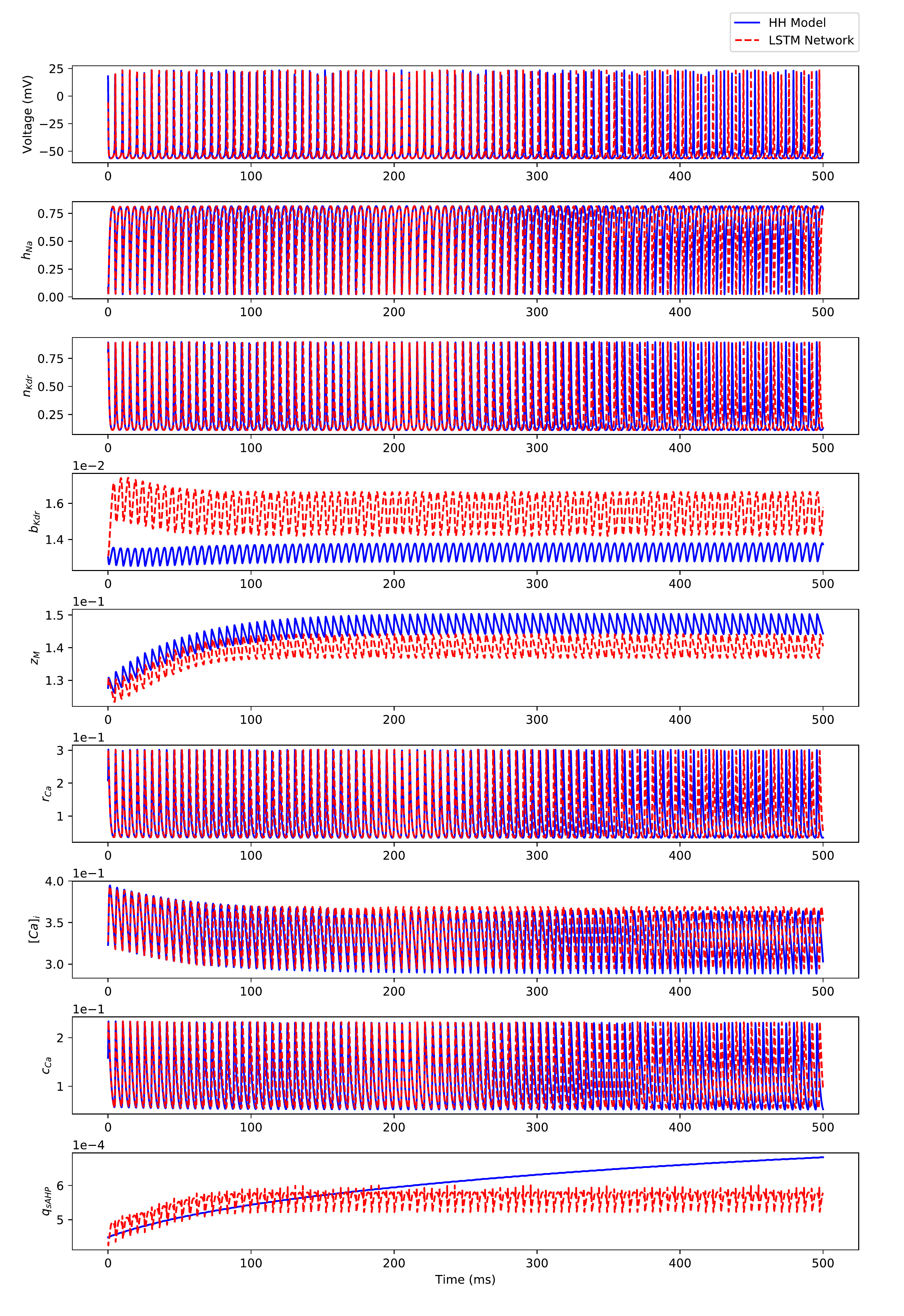}
	\caption{Comparison between the Hodgkin-Huxley model (``HH Model'') states' dynamics and the iterative predictions of states' dynamics using the 1 timestep predictive horizon-based deep LSTM neural network (``LSTM Network'')  in response to $I=3.0$ nA.}
	\label{fig:SS_Full-State_I=3,0}
\end{figure}
\begin{figure}[htp]
	\centering
	\includegraphics[scale = 0.5]{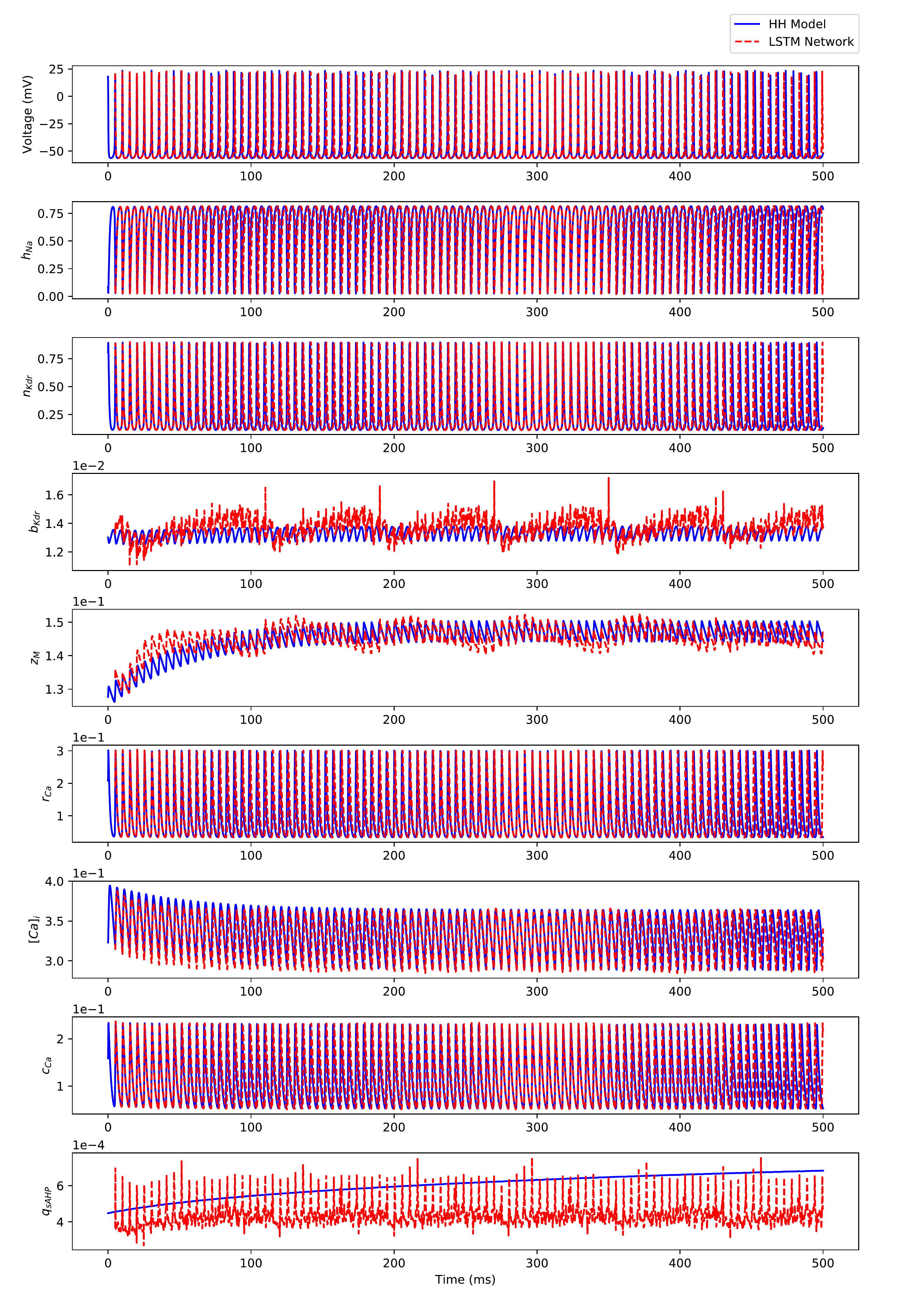}
	\caption{Comparison between the Hodgkin-Huxley model (``HH Model'') states' dynamics and the iterative predictions of states' dynamics using the 50 timesteps predictive horizon-based deep LSTM neural network (``LSTM Network'')  in response to $I=3.0$ nA.}
	\label{fig:50S_Full-State_I=3,0}
\end{figure}

\begin{figure}[htp]
	\centering
	\includegraphics[scale = 0.5]{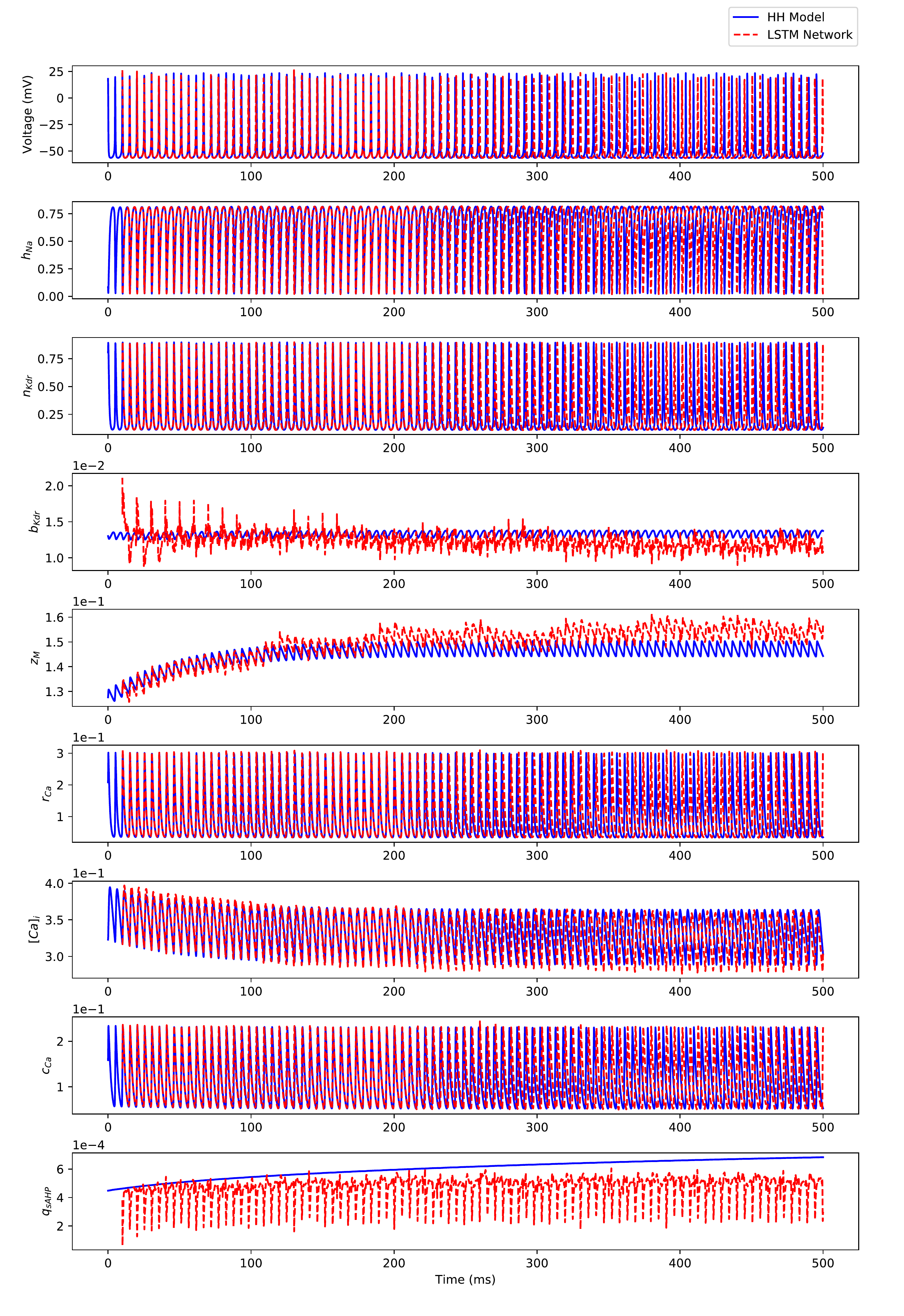}
	\caption{Comparison between the Hodgkin-Huxley model (``HH Model'') states' dynamics and the iterative predictions of states' dynamics using the 100 timesteps predictive horizon-based deep LSTM neural network (``LSTM Network'')  in response to $I=3.0$ nA.}
	\label{fig:100S_Full-State_I=3,0}
\end{figure}

\begin{figure}[htp]
	\centering
	\includegraphics[scale = 0.5]{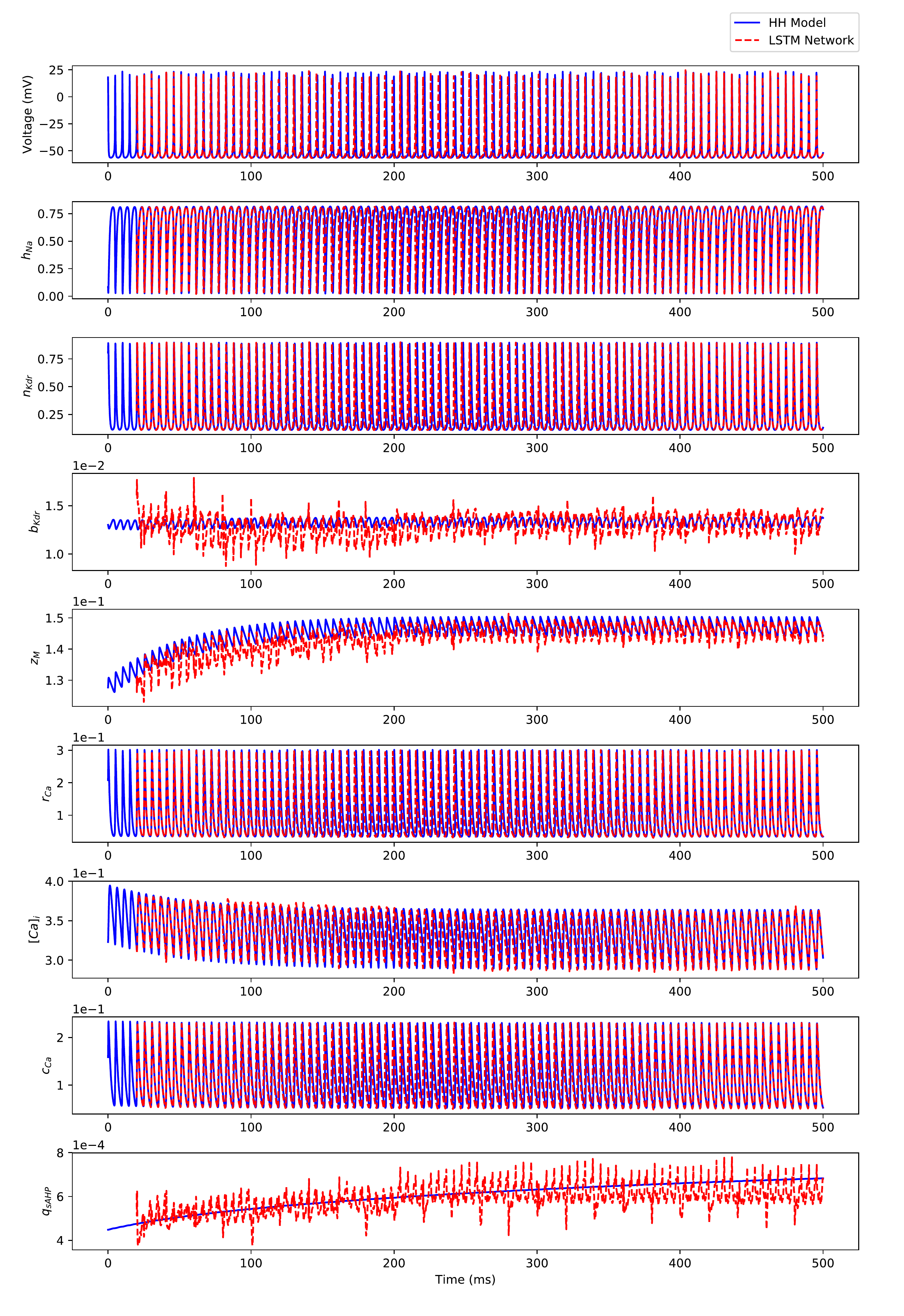}
	\caption{Comparison between the Hodgkin-Huxley model (``HH Model'') states' dynamics and the iterative predictions of states' dynamics using the 200 timesteps predictive horizon-based deep LSTM neural network (``LSTM Network'')  in response to $I=3.0$ nA.}
	\label{fig:200S_Full-State_I=3,0}
\end{figure}

\begin{figure}[htp]
	\centering
	\includegraphics[scale = 0.6]{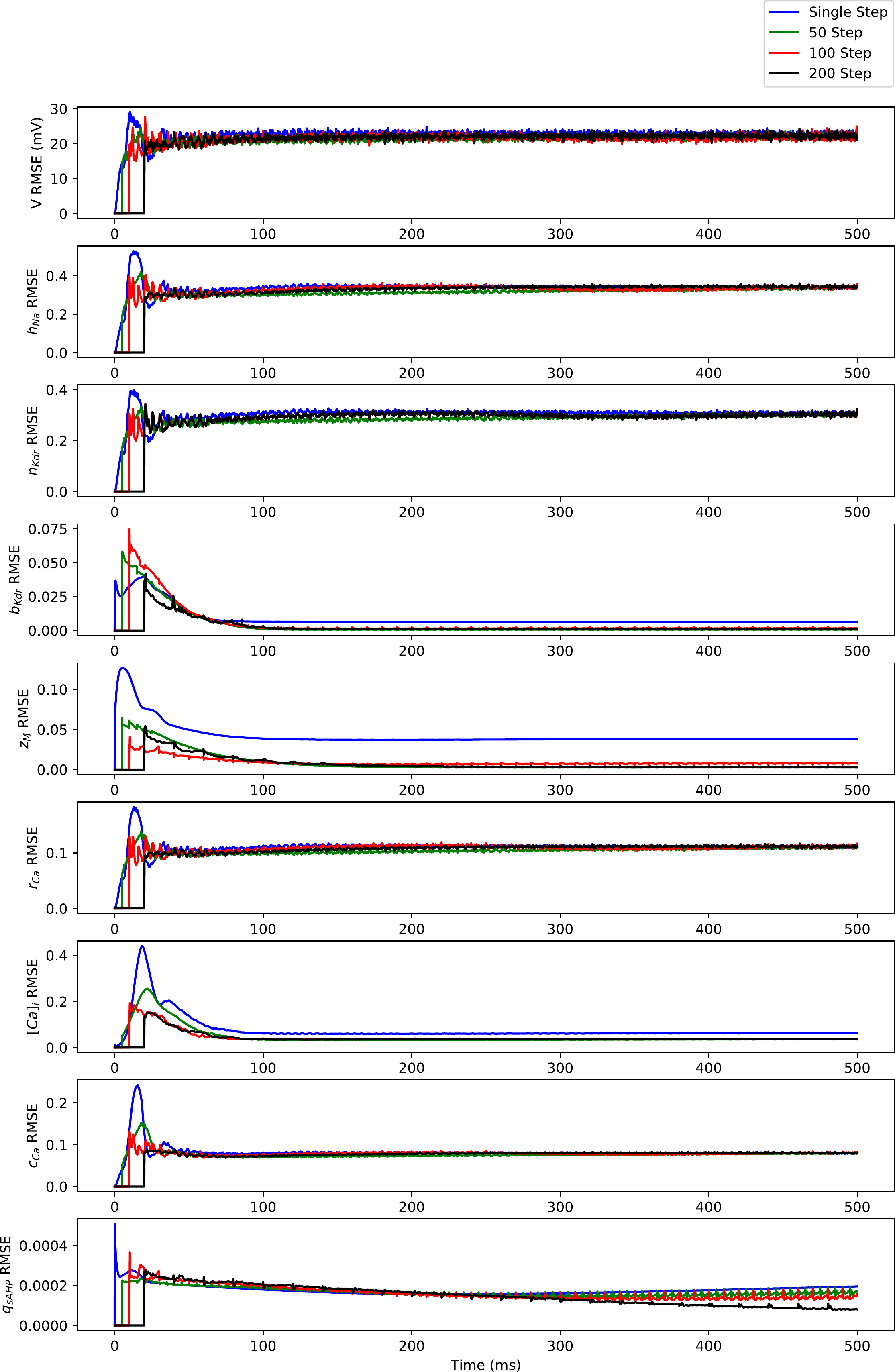}
	\caption{The root mean squared error (RMSE) versus simulation time for 5000 independent realizations, drawn from the predicted membrane potential trajectories of 50 randomly selected stimulating currents from a Uniform distribution $\mathcal{U}(2.3,3.0)$ and 100 random initial conditions for each stimulating current.}
	\label{fig:RMSEvsTimeHighFull}
\end{figure}

As shown in these figures, the performance of the deep LSTM neural network model in predicting state dynamics significantly improved with the increased predictive horizon of the LSTM network (i.e., $N_{p} = 1$ to $N_{p} = 200$) for all the states except $q_{sAHP}$ for which we found that the magnitude is comparable to the numerical precision of the performed simulations. Figure \ref{fig:RMSEvsTimeHighFull} shows the root mean squared error between the states of HHCA1Py and the deep LSTM neural network model as a function of simulation time over $5000$ random realizations, for $N_{p} = 1, 50, 100, 200$. These results show that the root mean squared error decreases from $N_{p} = 1$ to $N_{p} = 200$.  
\subsection{Irregular Bursting}\label{Appendix_IBS}
\begin{figure}[htp]
	\centering
	\includegraphics[scale = 0.5]{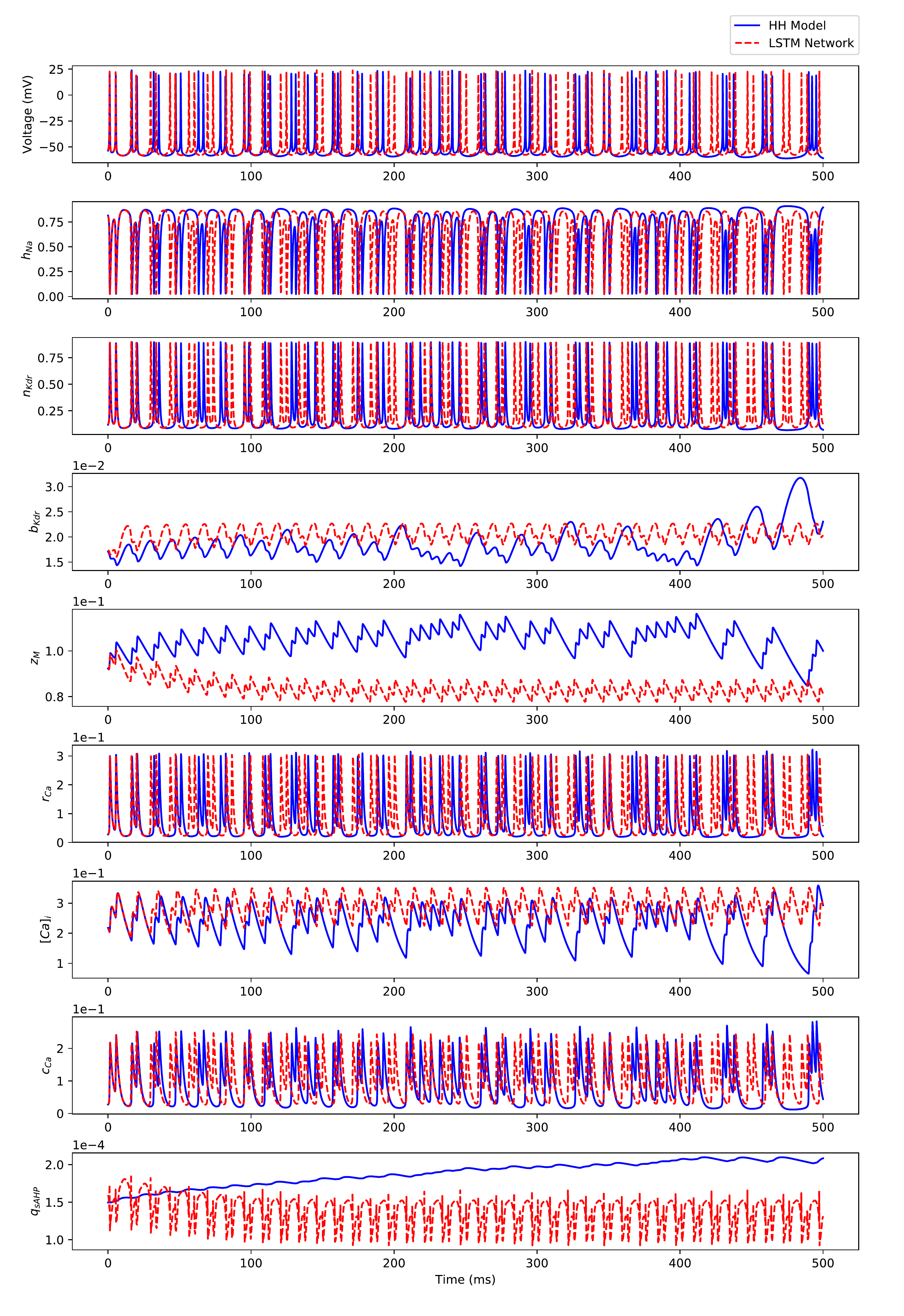}
	\caption{Comparison between the Hodgkin-Huxley model (``HH Model'') states' dynamics and the iterative predictions of states' dynamics using the 1 timestep predictive horizon-based deep LSTM neural network (``LSTM Network'')  in response to $I=1.5$ nA.}
	\label{fig:SS_Full-State_I=1.5}
\end{figure}

\begin{figure}[htp]
	\centering
	\includegraphics[scale = 0.5]{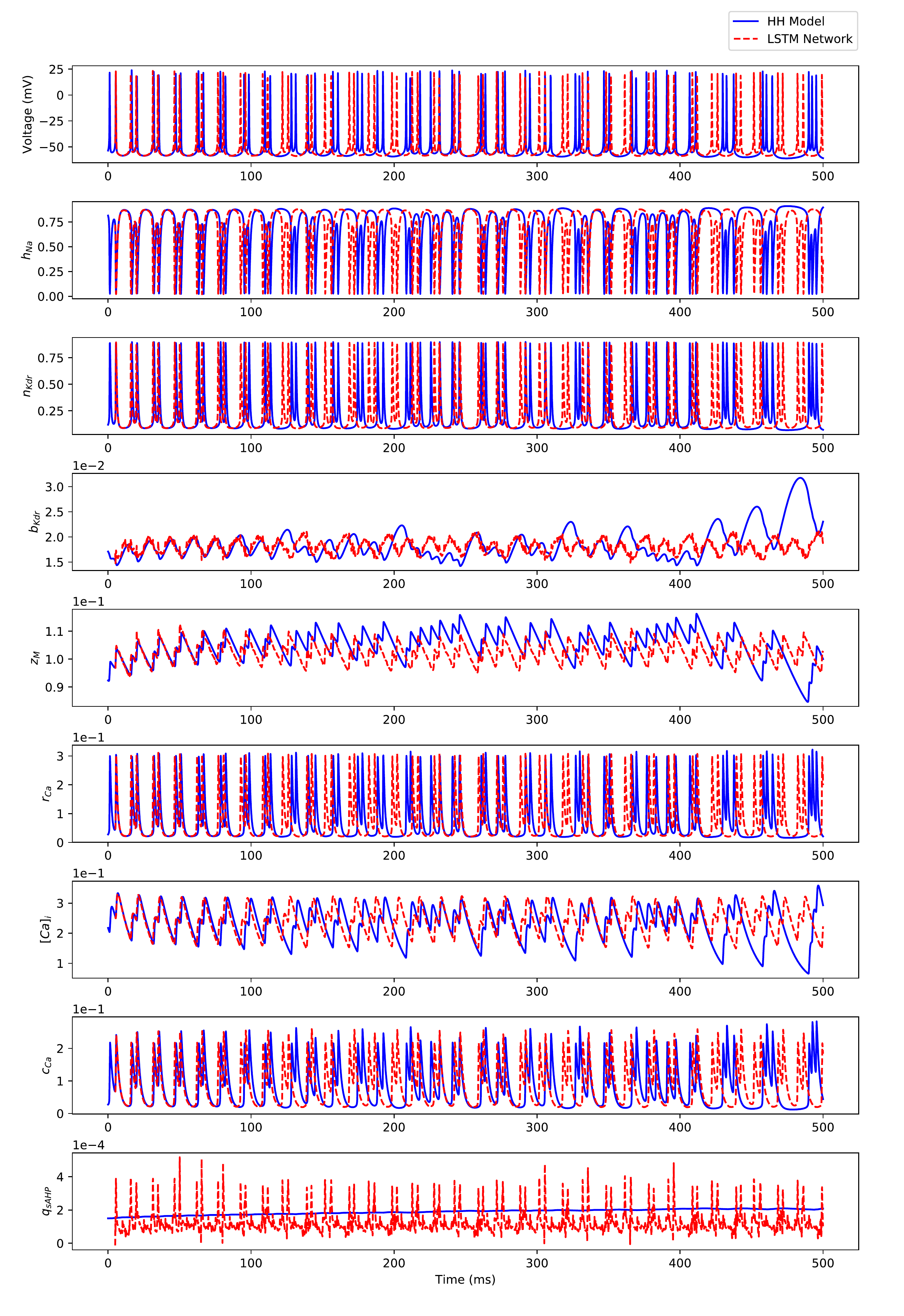}
	\caption{Comparison between the Hodgkin-Huxley model (``HH Model'') states' dynamics and the iterative predictions of states' dynamics using the 50 timesteps predictive horizon-based deep LSTM neural network (``LSTM Network'')  in response to $I=1.5$ nA.}
	\label{fig:50S_Full-State_I=1.5}
\end{figure}

\begin{figure}[htp]
	\centering
	\includegraphics[scale = 0.5]{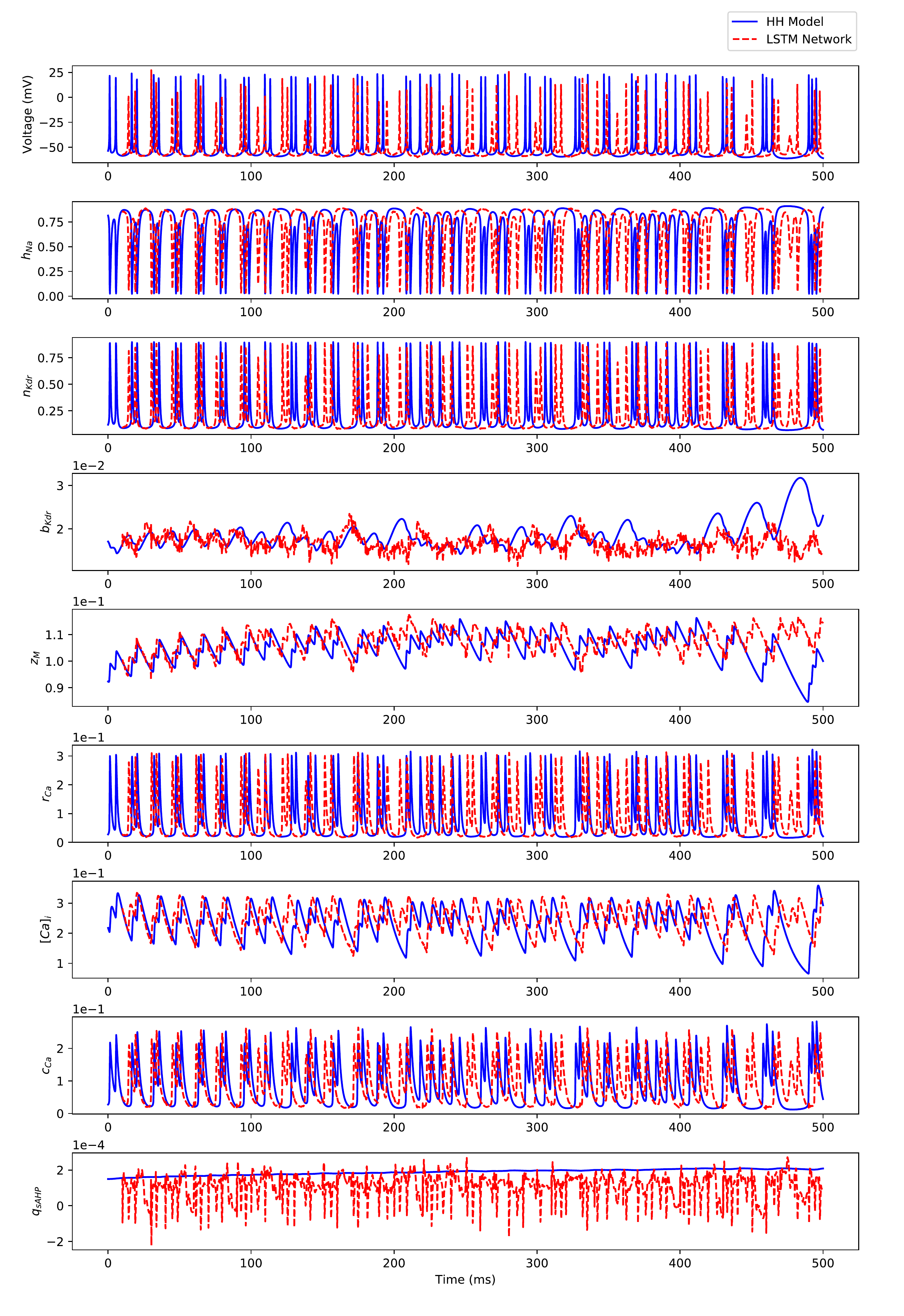}
	\caption{Comparison between the Hodgkin-Huxley model (``HH Model'') states' dynamics and the iterative predictions of states' dynamics using the 100 timesteps predictive horizon-based deep LSTM neural network (``LSTM Network'')  in response to $I=1.5$ nA.}
	\label{fig:100S_Full-State_I=1.5}
\end{figure}

\begin{figure}[htp]
	\centering
	\includegraphics[scale = 0.5]{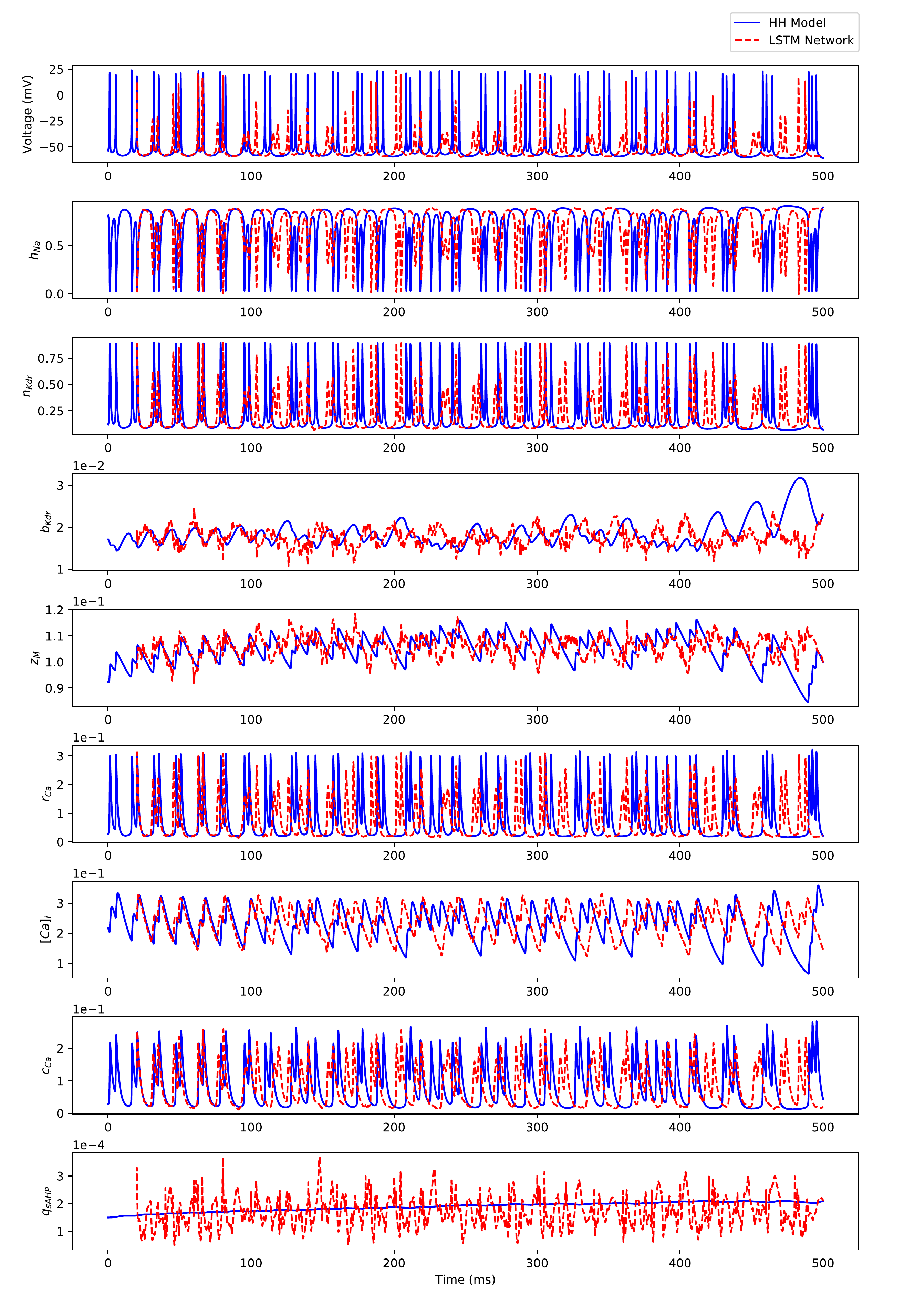}
	\caption{Comparison between the Hodgkin-Huxley model (``HH Model'') states' dynamics and the iterative predictions of states' dynamics using the 200 timesteps predictive horizon-based deep LSTM neural network (``LSTM Network'')  in response to $I=1.5$ nA.}
	\label{fig:200S_Full-State_I=1.5}
\end{figure}

\noindent In this section, we show the simulation results on predicting the dynamics of all the 9 states of HHCA1Py over a long-time horizon using the deep LSTM neural network for the irregular bursting regime ($I\in[0.79,2.3)$ nA).  Figures \ref{fig:SS_Full-State_I=1.5}, \ref {fig:50S_Full-State_I=1.5}, \ref {fig:100S_Full-State_I=1.5}, and \ref {fig:200S_Full-State_I=1.5} show the comparison between the state's dynamics simulated using the Hodgkin-Huxley model and the deep LSTM neural network model developed for 1 timestep, 50 timesteps, 100 timesteps, and 200 timesteps (equivalently, $N_{p} = 1,50,100,200$) predictive horizon, respectively.

\begin{figure}[htp]
	\centering
	\includegraphics[scale = 0.6]{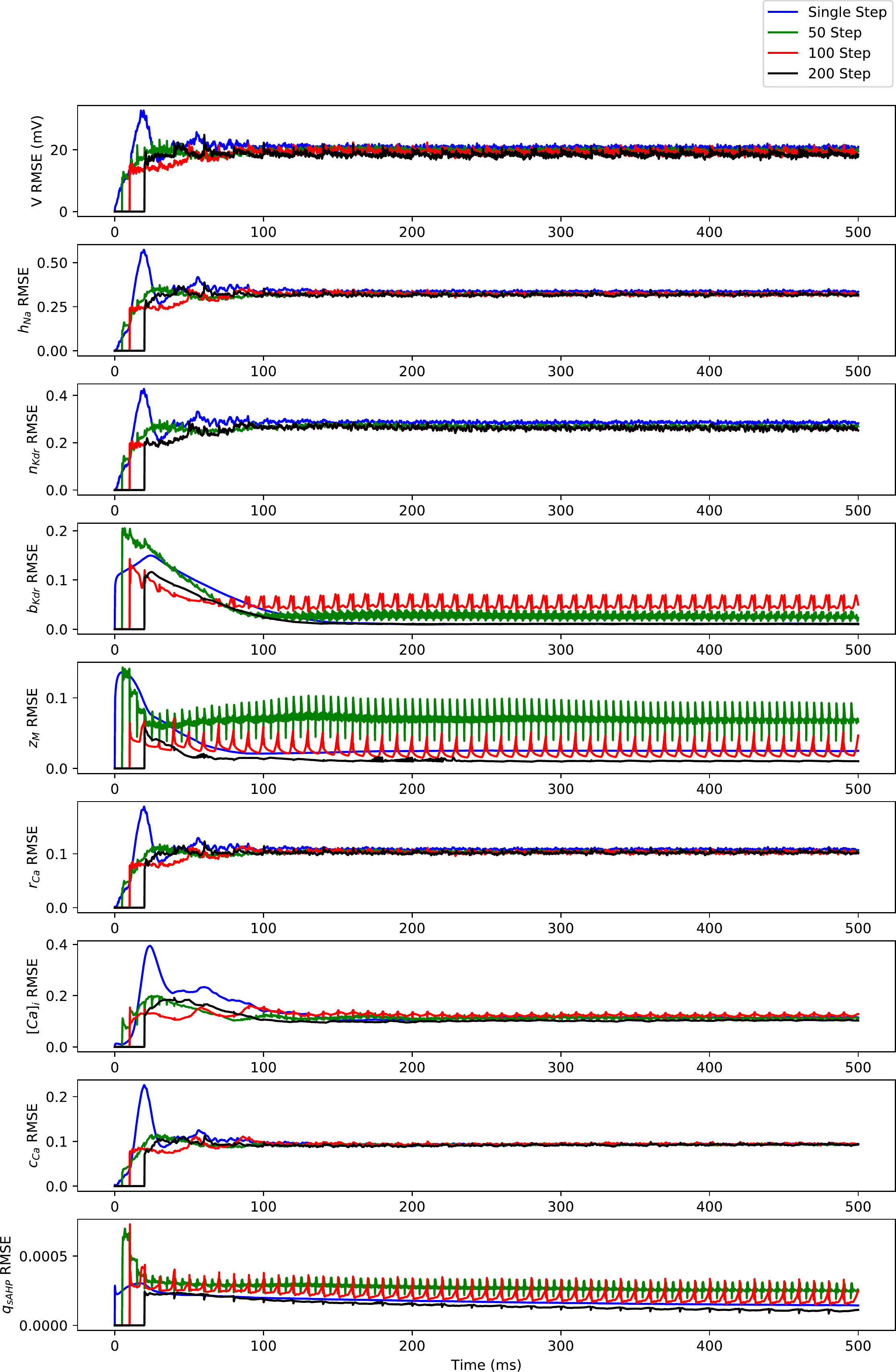}
	\caption{The root mean squared error (RMSE) versus simulation time for 5000 independent realizations, drawn from the predicted membrane potential trajectories of 50 randomly selected stimulating currents from a Uniform distribution $\mathcal{U}(0.79,2.3)$ and 100 random initial conditions for each stimulating current.}
	\label{fig:RMSEvsTimeMediumFull}
\end{figure}

As shown in these figures, the deep LSTM neural network model provides a reasonable prediction of the dynamics of all the states except $q_{sAHP}$ over the initial 100 ms of simulations. Moreover, the prediction improved from $N_{p} = 1$ to $N_{p} = 200$, which is consistent with the results for the regular spiking regime (see Figures \ref{fig:SS_Full-State_I=3,0}, \ref {fig:50S_Full-State_I=3,0}, \ref {fig:100S_Full-State_I=3,0}, \ref {fig:200S_Full-State_I=3,0}, and \ref {fig:RMSEvsTimeHighFull}). We found that the magnitude of $q_{sAHP}$ was comparable to the numerical precision of our simulations, which hindered the capability of the LSTM network in making a reasonable prediction for this state.

Figure \ref{fig:RMSEvsTimeMediumFull} shows the root mean squared error between the states of HHCA1Py and the deep LSTM neural network model as a function of simulation time over $5000$ random realizations, for $N_{p} = 1, 50, 100, 200 $. As shown here, the root mean squared error decreased with the increased predictive horizon of the LSTM network (i.e., $N_{p} = 1$ to $N_{p} = 200$).

\subsection{Regular Bursting} \label{Appendix_RBS}
\begin{figure}[htp]
	\centering
	\includegraphics[scale = 0.5]{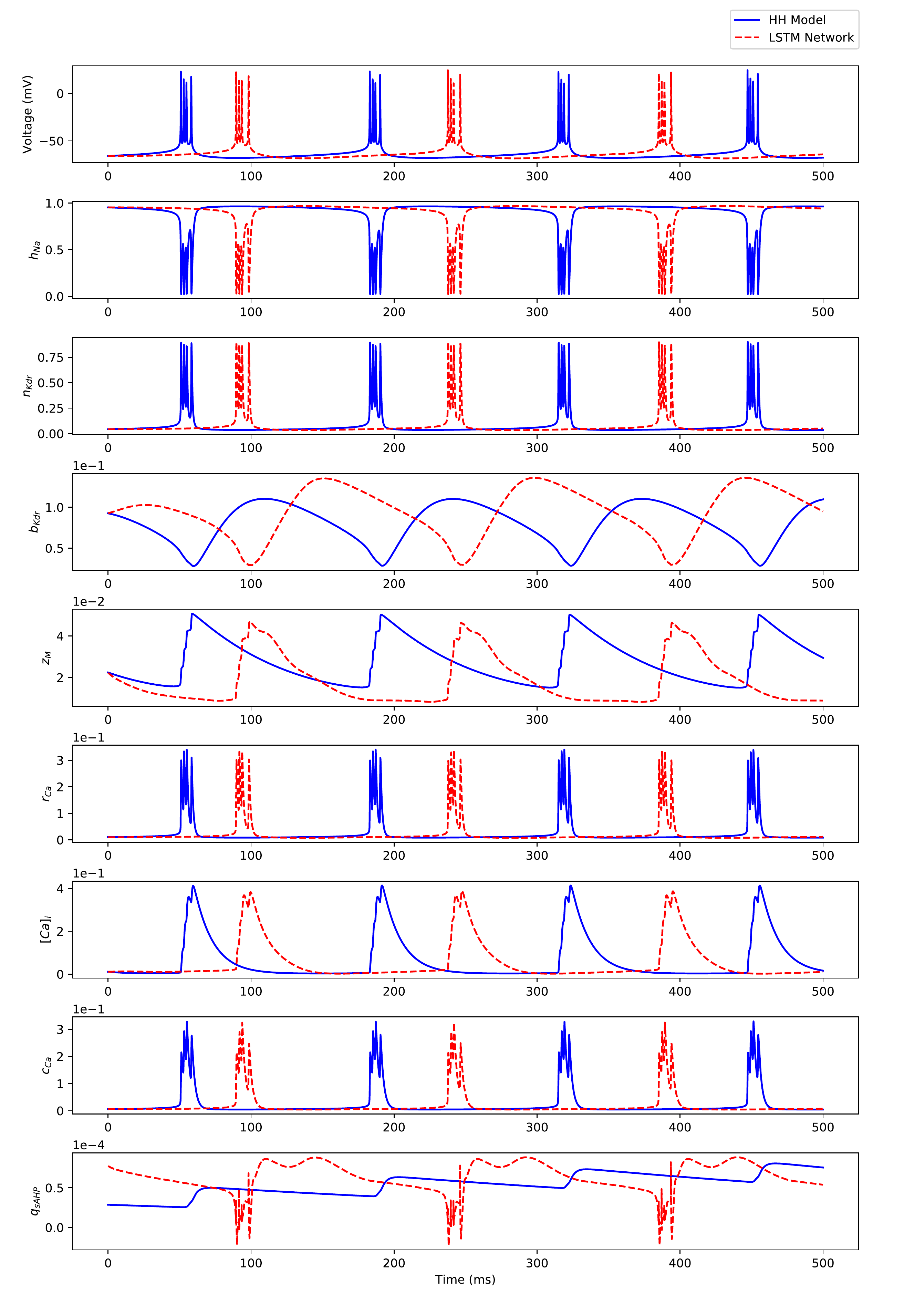}
	\caption{Comparison between the Hodgkin-Huxley model (``HH Model'') states' dynamics and the iterative predictions of states' dynamics using the 1 timestep predictive horizon-based deep LSTM neural network (``LSTM Network'')  in response to $I=0.5$ nA.}
	\label{fig:SS_Full-State_I=0.5}
\end{figure}

\begin{figure}[htp]
	\centering
	\includegraphics[scale = 0.5]{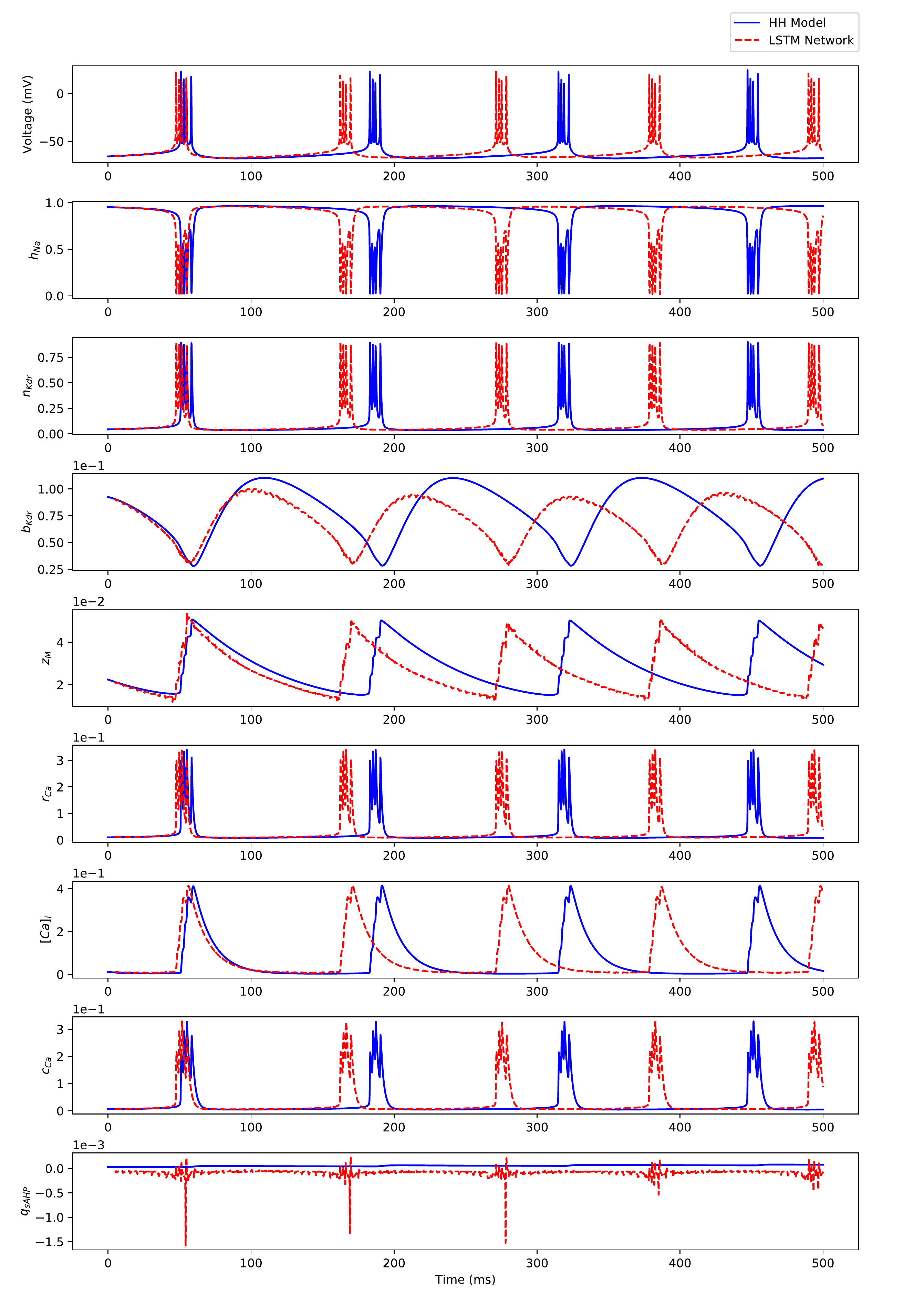}
	\caption{Comparison between the Hodgkin-Huxley model (``HH Model'') states' dynamics and the iterative predictions of states' dynamics using the 50 timesteps predictive horizon-based deep LSTM neural network (``LSTM Network'')  in response to $I=0.5$ nA.}
	\label{fig:50S_Full-State_I=0.5}
\end{figure}

\begin{figure}[htp]
	\centering
	\includegraphics[scale = 0.5]{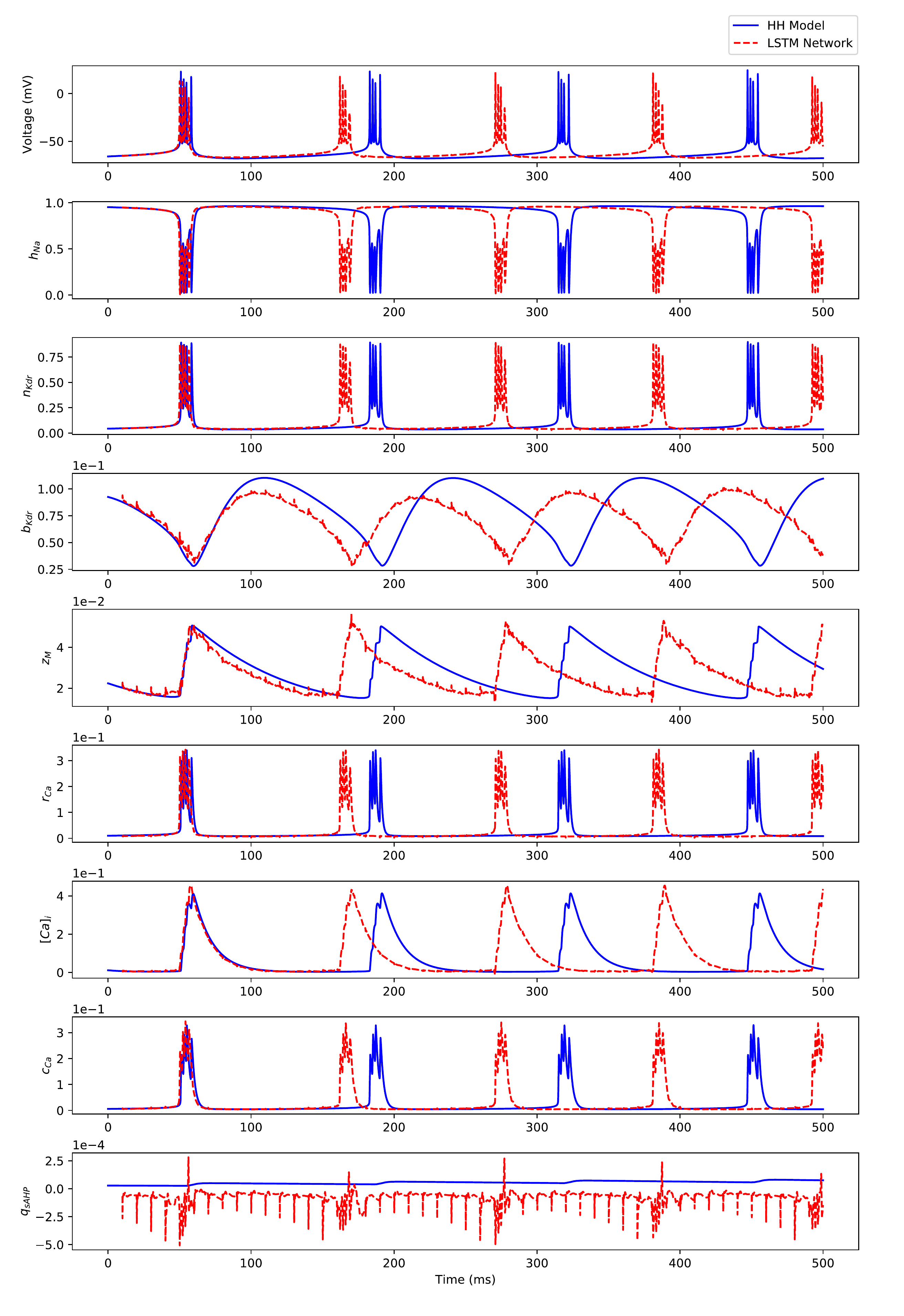}
	\caption{Comparison between the Hodgkin-Huxley model (``HH Model'') states' dynamics and the iterative predictions of states' dynamics using the 100 timesteps predictive horizon-based deep LSTM neural network (``LSTM Network'')  in response to $I=0.5$ nA.}
	\label{fig:100S_Full-State_I=0.5}
\end{figure}

\begin{figure}[htp]
	\centering
	\includegraphics[scale = 0.5]{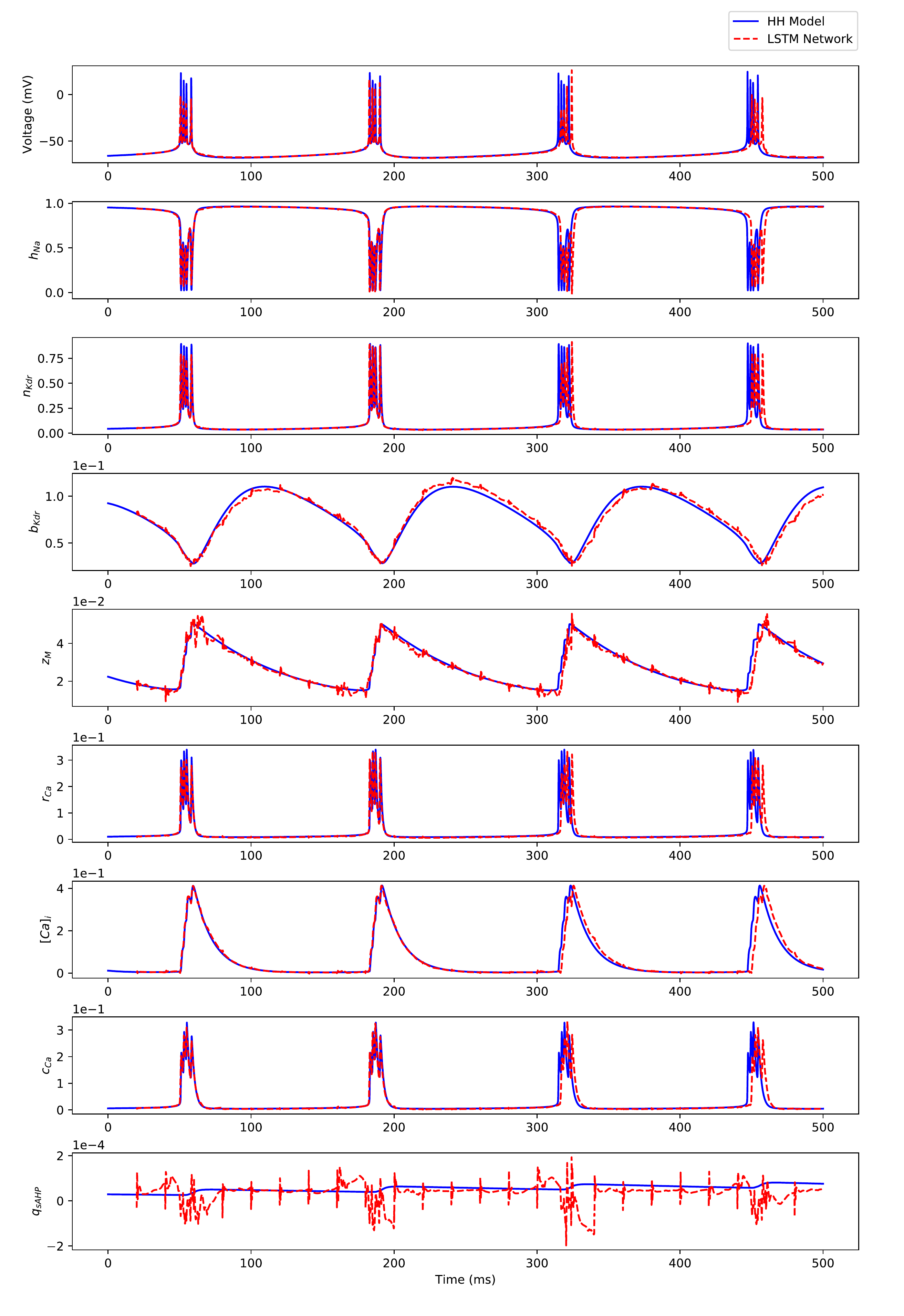}
	\caption{Comparison between the Hodgkin-Huxley model (``HH Model'') states' dynamics and the iterative predictions of states' dynamics using the 200 timesteps predictive horizon-based deep LSTM neural network (``LSTM Network'')  in response to $I=0.5$ nA.}
	\label{fig:200S_Full-State_I=0.5}
\end{figure}

\noindent In this section, we show the simulation results on predicting the dynamics of all the 9 states of HHCA1Py over a long-time horizon using the deep LSTM neural network for the regular bursting regime ($I\in[0.24,0.79)$ nA).  Figures \ref{fig:SS_Full-State_I=0.5}, \ref {fig:50S_Full-State_I=0.5}, \ref {fig:100S_Full-State_I=0.5}, and \ref {fig:200S_Full-State_I=0.5} show the comparison between the state's dynamics simulated using the Hodgkin-Huxley model and the deep LSTM neural network model developed for 1 timestep, 50 timesteps, 100 timesteps, and 200 timesteps (equivalently, $N_{p} = 1,50,100,200$) predictive horizon, respectively.

\begin{figure}[htp]
	\centering
	\includegraphics[scale = 0.6]{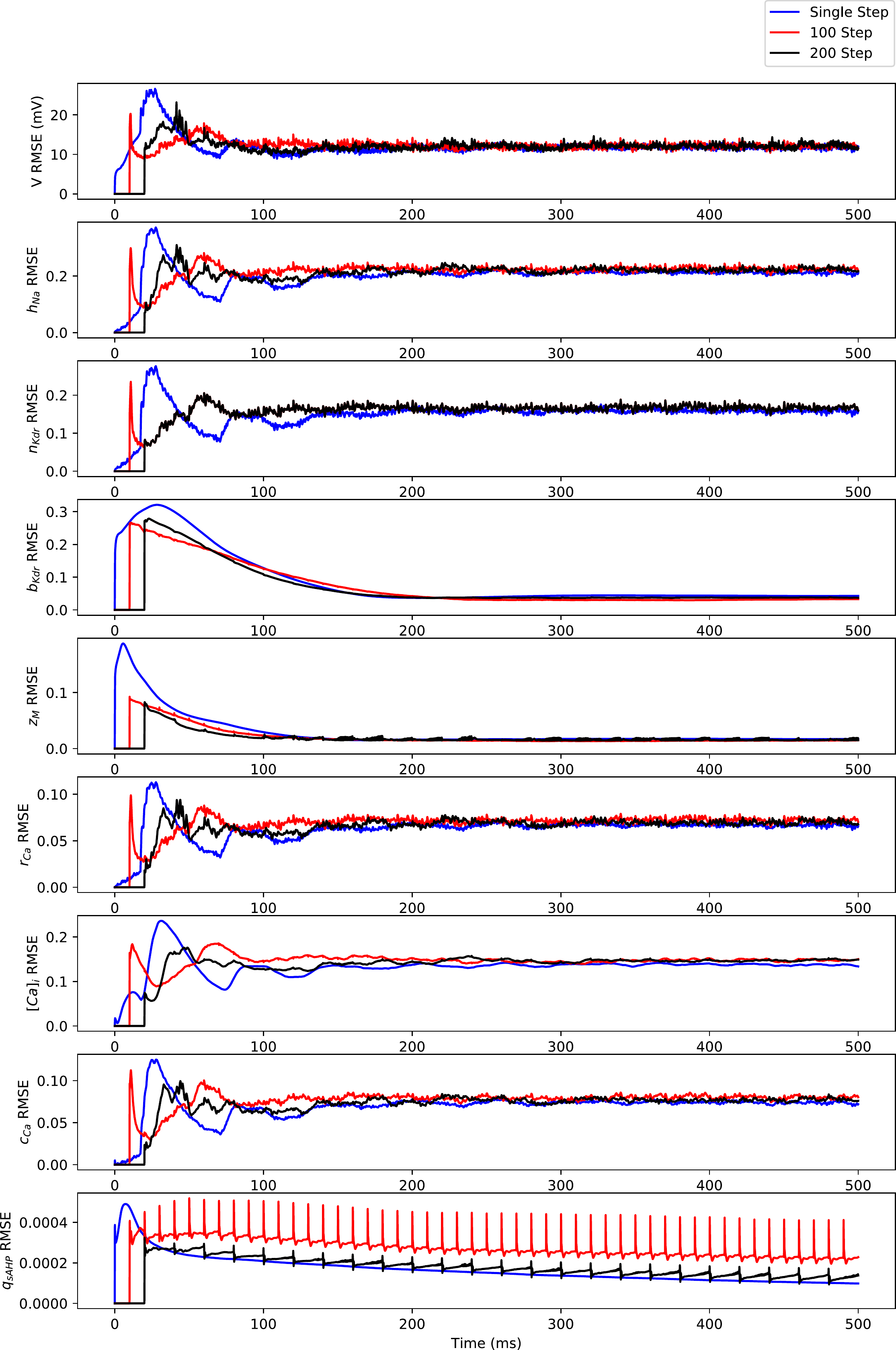}
	\caption{The root mean squared error (RMSE) versus simulation time for 5000 independent realizations, drawn from the predicted membrane potential trajectories of 50 randomly selected stimulating currents from a Uniform distribution $\mathcal{U}(0.24,0.79)$ and 100 random initial conditions for each stimulating current.}
	\label{fig:RMSEvsTimeLowFull}
\end{figure}

As shown in these figures, the performance of the deep LSTM neural network model in predicting state dynamics significantly improved between 1 timestep predictive horizon (Figure \ref{fig:SS_Full-State_I=0.5}) and 200 timesteps predictive horizon (Figure \ref {fig:200S_Full-State_I=0.5}) across all the states except $q_{sAHP}$ for the similar reason we provided for the regular spiking and irregular bursting regimes. More importantly, the LTSM network predicted the temporal correlations with high accuracy over the time-horizon of 300 ms for $N_{p} = 200$.  The extrapolation of these results suggest that increasing the predictive horizon beyond $N_{p} = 200$ could improve the prediction beyond 300 ms of time-horizon.

In Figure \ref{fig:RMSEvsTimeMediumFull}, we show the root mean squared error between the states of HHCA1Py and the deep LSTM neural network model as a function of simulation time over $5000$ random realizations, for $N_{p} = 1, 50, 100, 200 $. As shown here, the root mean squared error decreased with the increased predictive horizon of the LSTM network (i.e., $N_{p} = 1$ to $N_{p} = 200$), which is consistent with the results of the regular spiking and irregular bursting regimes.

\end{document}